\documentclass{cjour}
\usepackage{amsmath,amssymb,eucal,epsfig,latexsym}
\def\cY{{\cal Y}}
\def\cX{{\cal X}}
\def\cZ{{\cal Z}}

\def\cT{{\cal T}}
\def\cB{{\cal B}}

\def\cF{{\cal F}}
\def\cN{{\cal N}}

\def\cM{{\cal M}}

\def\cP{{\cal P}}
\def\cG{{\cal G}}
\def\cS{{\cal S}}

\def\eN{{\cal O}}

\def\eN{{\mathfrak N}}

\def\Blackboardfont{\mathbb}

\def\N{{\Blackboardfont N}}
\def\R{{\Blackboardfont R}}
\newcommand\tcomp{\mathfrak{T}}

\newcommand\rwr{\leadsto}
\newcommand\rewr[1]{\stackrel{#1}{\rwr }}
\newcommand\move[1]{\stackrel{#1}{\longrightarrow }}
\newcommand\mover[1]{\stackrel{#1}{\longleftarrow }}
\newcommand\moven[1]{\stackrel{#1}{\leftharpoonup\!\!\rightharpoonup }}
\newcommand\moved[1]{\stackrel{#1}{\rightharpoondown }}
\newcommand\moveu[1]{\stackrel{#1}{\rightharpoonup }}

\def\eref#1{(\ref{#1})}

\newcommand\mleq{ \leqslant }
\newcommand\mgeq{ \geqslant }

\newcommand\place{{\cal P}}
\newcommand\trans{{\cal T}}
\newcommand\transf{\varphi}
\newcommand\pn{{\cal N}}

\begin{document}
\thispagestyle{empty}
{\large
\parindent 0pt
\parskip6pt
\vspace*{1.5in}










\title{Blocking a  transition in a Free Choice net
and what it tells about its throughput\thanks{This work was 
supported by the European Community Framework IV programme through the
research network {\sc ALAPEDES} 
  (``The {\sc AL}gebraic {\sc A}pproach to {\sc P}erformance {\sc E}valuation
 of {\sc D}iscrete {\sc E}vent {\sc S}ystems''); S.H. was supported also by the project RNRT/MAGDA.
This work was started while S.H. was with
INRIA / \'{E}cole Normale Sup\'{e}rieure,                     
       Paris, France.}}

\author{Bruno Gaujal$^{\ast}$, Stefan Haar$^{\dagger}$, and 
Jean Mairesse$^{\ddagger}$
}

\affil{$^{\ast}$INRIA/ENS-Lyon, LIP, 46 All{\'e}e d'Italie,
69364 Lyon Cedex 07, France. Email:
bruno.gaujal@ens-lyon.fr; $^{\dagger}$INRIA/IRISA, Campus de
Beaulieu, 35000 Rennes, France.   
Email: Stefan.Haar@irisa.fr;
and $^{\ddagger}$CNRS-Universit{\'e} Paris 7, LIAFA, Case 7014,
2 place Jussieu, 75251 Paris Cedex 05,
France. Email: mairesse@liafa.jussieu.fr.} } 



\abstract{In a live and bounded Free Choice Petri net, pick
a non-conflicting transition. Then there exists a unique reachable
marking in which 
no transition is  
enabled except the selected one. 
For a routed live and bounded Free Choice net, this property is true for any 
transition of the net. 
Consider now a live and bounded stochastic routed Free Choice net, and assume
that the routings and the firing times are independent and identically
distributed.  
Using the above results,
we prove the existence of 
asymptotic firing throughputs for all transitions in the net. 
Furthermore, the vector of the throughputs at the different transitions is
explicitly computable up to a multiplicative constant.
} 

\begin{article} 

\section{Introduction}

The paper is made of three parts, each of which considers a different kind
of Petri nets. 
In the first part, we look at {\em classical} untimed Petri nets as
 studied in \cite{DeEs,mura}; more precisely, we study
  live and bounded {\em Free Choice nets (FCN)}.
Using standard Petri net techniques,
we show that, after blocking a non-conflicting
transition $b$,
there exists a unique reachable marking $M_b$ where no transition 
can fire but the blocked one. We call $M_b$  the {\em blocking
  marking} associated 
with $b$. Examples of Petri nets are given which satisfy any two of the
three properties (live,
bounded, free choice) and do not have a blocking marking. 

\medskip

In the second part, we look at {\it routed Petri nets},
where each place with several output transitions is equipped
with a routing function for the successive tokens entering the place.
More precisely, we consider live and bounded routed Free Choice nets with 
{\em equitable} routings. 
In this case, there exists a unique blocking marking for any transition, even
a conflicting one. Furthermore all the firing sequences avoiding the
blocked transition and  
leading to the blocking marking have the same Parikh vector (i.e., the
same letter content).  

\medskip

Introducing routings in a Petri net is, in some sense, an
impoverishment since it removes 
the non-determinacy in the evolution: routing resolves all conflicts. On the
other hand, it provides the 
right framework for an important enrichment of the model: the introduction
of time. 

\medskip

In the last section, we consider live and bounded timed routed Free
Choice nets in a stochastic setting.
We assume  the routings 
(at the places with several output transitions) to be random, and  the
firing of a transition to take some random amount of time. The
successive routings at a  
place and the successive firing times of 
a transition form sequences of i.i.d. r.v. (independent and
identically distributed random variables).
Using the so-called `monotone-separable framework' 
(see \cite{BaFo93b,BaMa95,bona}), we prove 
a {\em first order} limit theorem:
each transition in the net fires with an asymptotic
rate. 
The ratio between the rates at two different transitions is explicitly
computable and depends only 
on the routing probabilities and not on the firing times. 
At the end of Section \ref{sto}, we briefly discuss 
two types of extensions: {\em (i)}- first order results under stationary 
assumptions for the routings and the firing times; {\em (ii)}- second
order results, that is, the existence 
of a unique stationary regime for the marking process.

\medskip

We conclude the introduction 
by explaining the motivations for this study, which are two-fold. 
First, Free Choice Petri nets are an important subclass of Petri nets
which realize a good compromise between modelling power and the
existence of strong mathematical properties, as emphasized in
\cite{DeEs}. The existence of a blocking marking appears as a new and
fundamental property of FCN. It may turn out to be
helpful for instance in verification or in fault management, with the
blocking of a transition corresponding to
some breakdown in the system.

Second, this structural result enables us to study the asymptotic
behavior of $\textit{\em stochastic}$ FCN under {\em
  i.i.d. assumptions}. Stochastic Petri nets under markovian or
semi-markovian 
assumptions is a long standing domain of research, see for instance
\cite{ABDFC}. The aim for more generality, as well as some strong
evidence about the intrinsic complexity of the timed characteristics
in modern networks (such as the internet, see \cite{wwil}), suggest to
go beyond the markovian setting. In our context, it implies studying
stochastic Petri nets in which the sequence of firing times of a
transition is i.i.d. with a general distribution. Obviously, in such a
general setting, we can not expect to get explicitly computable
performance measures. Instead, we are glad to settle for qualitative
results about the existence of throughputs or stationary
regimes. This program was already carried out for several subclasses
of Petri nets: T-nets \cite{bacc92,BCOQ}, unbounded Single-Input Free
  Choice nets (a subclass of FCN) \cite{BaFG}, and
bounded and unbounded Jackson networks (a
subclass of Single-Input FCN) \cite{BaFo94,BaFM96}. 
Here, we complement the picture by considering bounded FCN with a
general topology, thus generalizing 
from the Jackson setting and allowing for synchronization and splitting 
of streams. At last, we should mention that the above program is
carried out in \cite{haas} for general Petri nets but assuming
that there exists a so-called regeneration point. Roughly
speaking, the results of this paper enable to prove the existence of
such a regeneration point for a large subclass of live and bounded
FCN, see Section \ref{sse-marking}. 

It might be appealing to go even beyond the i.i.d. framework by using
stationary assumptions instead. This would allow to account for the
dependence of the timed characteristics upon the period of the day or
of the year. 
For T-nets, Single-Input FCN, and Jackson networks, the analysis in the
above mentioned articles was performed under stationary
assumptions. 
We discuss the possibility of such an extension for live and bounded
FCN in Section \ref{sse-notiid}. 

\section{Preliminaries on Petri Nets}\label{se-prel}
\subsection{Basic definitions}
We use the notation $\N^*=\N\setminus\{0\}$ and $\R^*=\R\setminus\{0\}$. 
We denote by $x\mleq y$ the coordinate-wise ordering of $\R^k$, 
and write $x<y$ if $x\mleq y$ and $x\neq y$. 

\medskip

A {\em Petri net} is a 4-tuple $\pn = ({\cal P},{\cal T},{\cal F},
M)$, where $(\cP,\cT,\cF)$  
is a finite bipartite directed graph with set of nodes $\cP\cup\cT$, 
where $\cP\cap\cT=\emptyset$, and
set of arcs $\cF  
\subset ({\cal P}\times {\cal T})\cup ({\cal T}\times {\cal P})$, and where $M$ belongs to 
$\N^{\cP}$. To avoid trivial cases, we assume that the sets $\cP$ and
$\cT$ are non-empty.  
The elements of $\cP$ are called {\em places}, those of $\cT$ 
{\em transitions}; an element
of $\N^{\cP}$ is a {\em marking}, and $M$ is 
the {\em initial marking}. To emphasize the 
role of the initial 
marking, we sometimes denote the Petri net 
$\pn=({\cal P},{\cal T},{\cal F}, M)$ by $(\pn,M)$. 

\medskip

We apply the standard terminology of graph theory to Petri nets, 
and  assume throughout all
Petri nets considered to be connected  (without loss of generality). 

\medskip

A Petri net $\pn '=({\cP}',{\cT}',{\cF}', M')$ is a {\em subnet} of 
$\pn=({\cP},{\cT},{\cF}, M)$, written $\pn'=\pn[\cP'\cup \cT']$, if 
\[
\cP'\subset \cP, \cT'\subset \cT, \cF' =\cF \cap 
\left( \ (\cP'\times \cT')\cup (\cT'\times \cP') \ \right)\:,
\]
and $M'$ is the restriction of $M$ to $\cP'$.
If
$X\subseteq\cP\cup \cT$,
the {\em subnet generated by $X$} is the subnet $\pn[X]$.
We use the
classic graphical representation for Petri nets: circles for places,
rectangles for transitions,
and tokens for markings; see for example Figure \ref{fi-example}. 
 We write $x\to y$ if $(x,y)\in \cF$,
and denote by  
\[
^\bullet x = \{y \ : \  y\to x\}, \ \mbox{ and } \  
x^\bullet =\{y \ : \ x\to y\}\:,
\]
the sets of input/output nodes of a  node $x$.
The {\em incidence matrix}  $N\in \{-1,0,1\}^{\cP\times \cT}$ of $\pn$ is defined by 
$N(p,t) =1$ if $(t\to p,p\not\to t)$, $N(p,t) =-1$ if $(p\to t, t \not\to p)$, and $N(p,t)=0$ otherwise.

\medskip

Let $\cT^*$ be the free monoid over $\cT$, that is, the set of finite words 
over $\cT$ equipped with the concatenation product. We denote  the
empty word by $e$. 
Let $\cT^{\N}$ be the set of infinite words over the alphabet $\cT$.
Consider
a (finite or infinite) word $u$; we denote by 
$|u|$ its length (in $\N\cup\{\infty\}$)  and, for $a\in \cT$,  by $|u|_a$
the number of occurrences 
of $a$ in $u$. The prefix of length $k$ of $u$ ($k\in\N$, $k\mleq |u|$) is
denoted by  
$u_{[k]}$.  
Further, let $\vec{u}\in (\N\cup \{\infty\})^{\cT}$ denote
the {\em Parikh vector}
or {\em commutative image} of $u$, that is,
$\vec{u}=(|u|_a)_{a\in \cT}$. 

\medskip

In a Petri net, the marking evolves with the {\em firing} of transitions. 
A transition $a$ is {\em enabled} 
in the marking $M$ if for all  place $p$ in ${}^{\bullet}a$, $M(p)>0$;
an enabled transition $a$  
can {\em fire}; the {\em firing} of $a$ 
transforms the marking $M$ into $M'=M + N\cdot \vec{a}$, written $M\move{a} M'$.
We say that a
word $u\in \cT^*$ is a {\em firing sequence} of $(\pn,M)$ if for all
$k\mleq |u|$, we have  
$M + N\cdot \vec{u}_{[k]} \mgeq (0,\dots ,0)$;
we say that $u$ transforms $M$ into 
$M'=M + N\cdot \vec{u}$, in which case we write $M\move{u} M'$. An 
infinite word over
$\cT$ is an  
{\em infinite firing sequence} 
if all its prefixes are firing sequences. 
The notation $M\move{u}$ means that $u$ is a (infinite) firing
sequence of $(\pn,M)$.  
A marking $M_2$ is {\em reachable} from a marking $M_1$ if there
exists a firing sequence  
$u\in \cT^*$ such that 
$M_1\stackrel{u}{\longrightarrow} M_2$.
The set of {\em reachable markings} of $(\pn,M)$ is $R(\pn,M)= \{ M' \
: \ \exists u \in \cT^*, M \move{u}M'\}$. We write $R(M)$ instead of
$R(\pn,M)$  
when there is no risk of confusion.

\medskip

The Petri net $(\pn,M)$ is {\em live} if: $\forall M' \in R(M), \forall a \in \cT,
\exists M'' \in R(M'),  M''\move{a}$. A simple consequence of this
definition is that a live Petri net admits 
infinite firing sequences. 
The Petri net is {\em $k$-bounded} ($k\in\N$) if: $\forall M'\in R(M),
\forall p\in \cP, M'_p \mleq k$. The Petri net is {\em bounded} if it
is $k$-bounded for some $k\in \N$. 
A {\em deadlock} is a reachable marking in which no transition is enabled.

\medskip

A Petri net   $\pn=(\place , \trans, F,M)$ is a
\begin{itemize}
\item {\it T-net} (or {\em event graph}, or {\em marked graph}) if:  
$\forall p \in \place, \quad |{}^\bullet p| =| p^\bullet | = 1$;
\item {\it S-net} (or {\em state machine}) if: $\forall q \in \trans, \quad
  |{}^\bullet q| =| q^\bullet | = 1$; 
\item {\it Free Choice net}\footnote{see the remark on {\sf Extended
      Free Choice nets} in  Section \ref{se-fce}.}
(FCN) if: $\forall (p,q) \in \cF \cap (\cP\times \cT), \quad
p^{\bullet} =\{q\} \ \lor \ {}^{\bullet}q=\{p\}$.
\end{itemize}
An equivalent definition for a FCN is: $\forall q_1,q_2\in \trans, q_1\neq q_2, \
(p \in {}^\bullet q_1 \cap {}^\bullet q_2) \Rightarrow ( {}^\bullet q_1 =  {}^\bullet q_2  = \{ p\}).$
Obviously, every T-net is an FCN and  every S-net
is an FCN as well. 

In this paper, we study the class of live and bounded Free Choice nets. 
The membership of a given
Petri net to this class can be checked in polynomial time 
(in the size of the net), see 
for instance \cite{DeEs}, Chapter 6. 

\subsection{Additional background} \label{ap:results} 

This section can be skipped without too much
harm. Indeed, we gather the definitions and results to be needed in
the technical parts of different proofs (mainly the one of Theorem 
\ref{th:want}). 


\medskip

Proofs for the following results are given in \cite{DeEs}; for the
original references, 
see the bibliographic notes of \cite{DeEs}. 

\begin{theorem}[\cite{DeEs}, Theorem 2.25]\label{th:strongco}
A live and bounded connected Petri net is strongly connected.
\end{theorem}

A vector $X\in \N^{\cT}$ is a {\em T-invariant} if $N\cdot X =(0,\dots ,0)$.
If $u$ is a firing sequence 
such that $M\move{u} M$ then $\vec{u}$ is a T-invariant.

\begin{proposition}[\cite{DeEs}, Prop. 3.16] \label{le:tinvglatt}
In a connected T-net, the T-invariants are the vectors 
$(x,\dots ,x)$ for  $x\in \N$.
\end{proposition}

\begin{proposition}[\cite{mura}, Theorem 19]
\label{pr-new}
In a live T-net $(\pn,M)$ with incidence matrix $N$, if a vector $x\in
\N^{\cT}$ is such that $M+N\cdot x\mgeq (0,\dots ,0)$,
then there exists a firing sequence $u$ such that 
$\vec{u}=x$. 
\end{proposition}


\begin{proposition}[\cite{DeEs}, Theorem 3.18]\label{DE318}
A live T-net $(\pn,M)$ is 
$k$-bounded if and only if, for every place $p$, there exists a 
circuit which contains $p$ and holds at most $k$ tokens
under $M$.
\end{proposition}


A subnet $\pn'=(\cP',\cT',\cF',M')$ of $\pn$ is a {\em T-component}
(resp. {\em S-component}) if 
$\pn'$ is a strongly connected T-net (resp. S-net) and 
 satisfies: $\forall q \in \cT', \ ^\bullet q, q^\bullet \subseteq \cP'$ 
(resp. $\forall p \in \cP', \ ^\bullet p, p^\bullet\subseteq \cT'$). 
A set of subnets of $\pn$ forms a {\em covering} of $\pn$ if each node
and arc belongs to at least 
one of the subnets. 

\begin{theorem}[\cite{DeEs}, Theorems 5.6 and 5.18]
\label{th:TScover}
Live and bounded Free Choice nets are covered by S-components and by
T-components. 
\end{theorem}

The  {\it cluster} $[x]$ of a node $x$ in $\pn$ is the smallest  
subset of ${\place\cup \trans}$ such that \\
\hspace*{0.5cm} (i) $x\in[x]$; \ \ \
(ii) $p\in \place \cap [x] \ \Rightarrow \ p^\bullet \in \trans \cap [x]$; \ \ \ 
(iii) $q\in\trans \cap [x] \ \Rightarrow \ {}^\bullet q \in \place \cap [x]$.\\
If $\cG$ is a subnet of $\pn$, then the {\em cluster} $[\cG]$ of $\cG$ 
is the union of the clusters of all the nodes in $\cG$. 

\begin{theorem}[\cite{DeEs}, Theorem 5.20]
\label{th-acti}
Let $\cN'$ be a $T$-component of a live and bounded
Free Choice net $(\cN,M_0)$. 
There exists a firing sequence $\sigma$ containing no transition
from $[\cN']$ and such that $M_0\move{\sigma}M$ and $(\cN', M|_{\cN'})$
is live. 
\end{theorem}

Actually, Theorem 5.20 in \cite{DeEs} states that 
the sequence $\sigma$ does not contain any transitions from $\cN'$; 
however, the 
proof given in \cite{DeEs} also provides the result stated above (and this
strong version is the one we need).

\medskip

A {\it siphon} is a set of places $S$ such that
${}^\bullet S \subset S^\bullet $. A {\it trap} is a set of 
places $S$ such that $ S^\bullet\subset{}^\bullet S$.
In particular, if a siphon (resp. a trap) is
empty (resp. non-empty) under marking $M$, then it
remains empty (resp. non-empty) under all markings in $R(M)$.
The following theorem is known as Commoner's Theorem.

\begin{theorem}[\cite{DeEs}, Theorems 4.21  and 4.27]
\label{th:commoner}
A Free Choice net is live if and only if every siphon contains
an initially marked trap.
\end{theorem}

The fine structure of the dynamics in intersecting T-components leads
us to considering 
the subnets $\pn'$ such that any
given T-component 
either contains {\em no} or {\em all} transitions of  $\pn'$. These
are captured by the following definition.
A subnet $\pn'=(\place',\trans',F',M')$ of $\pn$
is a {\em CP-subnet} if (i) $\pn'$ is a non-empty and connected T-net; 
(ii) $\forall p\in \cP', \ ^\bullet p, p^\bullet \subseteq \cT'$;
(iii) the subnet generated by  
$(\place-\place')\cup (\trans-\trans')$ is strongly connected. 


A {\em way-in} (resp. {\em way-out}) transition of a Petri net is a
transition $a$ such that 
$^\bullet a=\emptyset$ (resp $a^\bullet=\emptyset$).

\begin{proposition}[\cite{DeEs}, Prop. 7.10]\label{pr:single} 
Let $\hat{\pn}$ be a CP-subnet of a live and bounded Free Choice net and let
$\hat{\trans}_{in}$ be
the set of way-in transitions of $\hat{\pn}$. We have 
$|\hat{\trans}_{in}|=1$.
\end{proposition}

\begin{proposition}[\cite{DeEs}, Prop. 7.8] \label{le:drumrum}
Let $(\pn,M_0)$ be a live and bounded  Free Choice net, let $\hat{\pn}$
be a CP-subnet of $\pn$ and let 
$\hat{\trans}$ be the set of transitions of $\hat{\pn}$ and $\hat{\trans}_{in}$ 
the set of way-in transitions of $\hat{\pn}$. 
Then there exists a marking $M$ and
a firing sequence $\sigma\in (\hat{\trans}-\hat{\trans}_{in})^*$ such that 
$M_0\move{\sigma}M$ and $M$ enables no transition of 
$\hat{\trans}-\hat{\trans}_{in}$.
Furthermore,
the subnet of $(\pn,M)$ generated by
$(\trans-\hat{\trans})\cup(\place-\hat{\place})$ 
is live and bounded. 
\end{proposition}



We now introduce the notion of reverse firings. 
Let $\cN$ be a Petri net. For a transition $q$ and two markings $M_1$
and $M_2$, we write 
\[
M_2\move{q^-}M_1 \ \mbox{ if } \ M_1\move{q}M_2\:.
\]
Given $u=u_1\cdots u_n, u_i \in \cT,$  we set $u^-=u_n^-\cdots u_1^-$. 
We write 
 $M_2\move{u^-} M_1$ if $M_1\move{u}M_2$.
We say that the {\em firing}
of $u^-$, or the {\em reverse firing} of $u$, transforms the marking
$M_2$ into $M_1$.  
Let us denote as $\trans^-=\{q^-:\
q\in\trans\}$ the set of {\em reverse transitions}. Given 
$u\in (\cT\cup\cT^-)^*$,  its {\em Parikh vector} is $\vec{u}=(|u|_a-|u|_{a^-})_{a\in \cT}$. 
A {\em generalized firing sequence} of
$(\pn,M)$ is a word $u\in (\cT\cup\cT^-)^*$ such that for all $k\mleq |u|$,
$M+N\cdot \vec{u}_{[k]} \mgeq (0,\dots ,0)$.

Define the following rewriting rules:
\begin{equation}\label{eq-rewr}
\forall a \in \cT, \ aa^- \rwr e, a^-a  \rwr e, \ \ 
\forall a,b \in \cT, a\neq b, \ ab^-  \rwr b^-a, b^-a  \rwr ab^-\:.
\end{equation}
For two words $u,v\in (\cT\cup\cT^-)^*$, 
we write $u\rewr{*}v$ if we can
obtain $v$ from $u$ by 
successive application
of a finite number of rewritings. 

\begin{lemma}\label{le:lemmaA}
Let $\pn$ be a T-net. Let $u,v \in (\cT\cup\cT^-)^*$ be such that
$u\rewr{*}v$. If $u$ is a 
generalized firing sequence,  
then $v$ is also a generalized firing sequence.
\end{lemma}

\begin{proof}
In a T-net, for two distinct
transitions $a$ and $b$, we have  
$a^\bullet \cap b^\bullet = \emptyset$ and $^\bullet a \cap ^\bullet b =
\emptyset$.  
The proof follows easily.
\end{proof}

\section{Blocking a Transition in a Free Choice net}


\subsection{Statement of the main result}

Let $(\pn,M)$ be a Petri net. 
A transition $a$ is a {\em non-conflicting} transition if for all $p
\in\ ^\bullet a, \ |p^\bullet|=1$;  
otherwise $a$ is a {\em conflicting} transition. 
We set $R_q(M)$ (resp. $R_q'(M)$) to be the set of markings reachable from $M$ 
(resp. reachable from $M$ without 
firing transition $q$) and in which no transition is
enabled except $q$:
\begin{eqnarray}
R_q(M)&=& \left\{ M': \ M'\in R(M), \ \left(\tilde{q}\in \cT, \ M' \move{\tilde{q}} \ \ 
    \Rightarrow \ \tilde{q}=q  \right) \right\}  
\label{eq-rb} \\
R_q'(M)&=& \left\{ M': \ M'\in R_q(M), \ \exists \sigma \in (\cT - \{q\})^*, 
M\move{\sigma} M'
\right\}\:.\nonumber 
\end{eqnarray}
As previously, we extend the notation to  $R_q(\cN,M)$ (resp. $R_q'(\cN,M)$)
when there is a possibility for ambiguity.  


\medskip

The next theorem is the heart of the article.

\begin{theorem}[Blocking one transition]
\label{th:want}
Let $(\pn,M_0)$ be a live and bounded 
Free Choice net. If $b$ is a  non-conflicting transition, then there exists 
a unique reachable marking $M_b$ in which the only enabled transition is $b$.
Furthermore, $M_b$ 
can be reached from any reachable marking and without firing transition $b$. 
\end{theorem}

Using the above notations, the result can be 
rephrased as: 
\[
\forall M\in R(M_0), \ R_b(M)=R_b'(M)=\{M_b\}\:.
\]

We call $M_b$ the {\em blocking marking} associated
with $b$. 
Note that a blocking marking is a {\em home state}, meaning that it is
reachable from any reachable marking.  

\begin{example}\label{ex-1}
To illustrate Theorem \ref{th:want}, 
consider the live  and bounded Free Choice net represented on the left
of Figure \ref{fi-example}.  The blocking markings associated with the
three non-conflicting transitions  
have been represented on the right of the figure. 
\begin{figure}[htb]
\begin{center}
\input{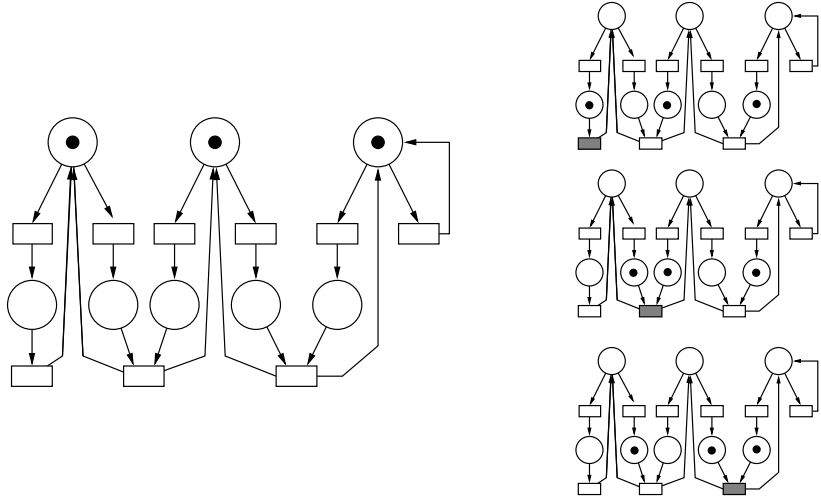}
\caption{Blocking markings associated with the non-conflicting
  transitions.\label{fi-example}} 
\end{center}
\end{figure}
\end{example}

Now the natural question is: 
do there always exist non-conflicting transitions? The answer is 
given in the next lemma.

\begin{lemma} Let $\cN$ be a live and bounded Free Choice net. 
If $\cN$ is not an S-net, then it contains non-conflicting transitions. 
\end{lemma}

\begin{proof}
The net $\cN$ is
strongly connected (Theorem \ref{th:strongco}), hence each node has at
least one predecessor and one successor.  
Due to the Free Choice property, a sufficient condition for a
transition $a$ to be   
non-conflicting is that $|^\bullet a|>1$. Assume that all transitions $a$
are such that  
$|^\bullet a|=1$. Since $\cN$ is not an S-net, there exists at least one
transition $t$ such that 
$|t^\bullet |>1$. 
If we have  $M \move{a} M', \ a\in \cT,$
then $\sum_p M_p'=\sum_p M_p +|a^\bullet| - |^\bullet a|$.  
Since $|^\bullet a|=1$ for all $a$ in $\cT$, the total number 
of tokens never decreases. On the other hand, if we have $M \move{t} M'$, 
then $\sum_p M_p'\mgeq  \sum_p M_p  +1$. Since the net is live,
 there exists an
infinite firing sequence $\sigma \in \cT^\N$ such  
that $t$ occurs an infinite number of times in $\sigma$. We deduce that
the total number of tokens along the markings 
reached by $\sigma$ is unbounded. This is a contradiction.
\end{proof}

\begin{figure}[htbp]
 \begin{center}
\input{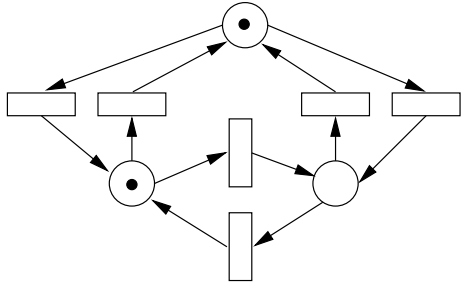}
    \caption{A live and bounded $S$-net without any non-conflicting
      transition.\label{fig:noblocking}} 
  \end{center}
\end{figure}

On the other hand, it is possible for an S-net to contain only
conflicting transitions.  
An example is displayed in Figure \ref{fig:noblocking}; there exists
no marking in which only one transition  
is enabled.

However, in all cases, if one blocks a cluster (see Section
\ref{ap:results}) instead of a single
transition,
then the net reaches a unique marking, the {\em blocking
marking associated with the cluster}. 

\begin{corollary}\label{co-cluster}
Let $\cN$ be a live and bounded Free Choice net. 
Let $b$ be any transition of $\cN$ and let $[b]$ be the cluster of $b$.
There exists 
a unique reachable marking $M_{[b]}$ in which the set of enabled
transitions is exactly the set of transitions in $[b]$.
Furthermore, the marking $M_{[b]}$ 
can be reached from any reachable marking and without firing any
transitions in $[b]$. 
\end{corollary}

\begin{figure}
  \begin{center}
  \input{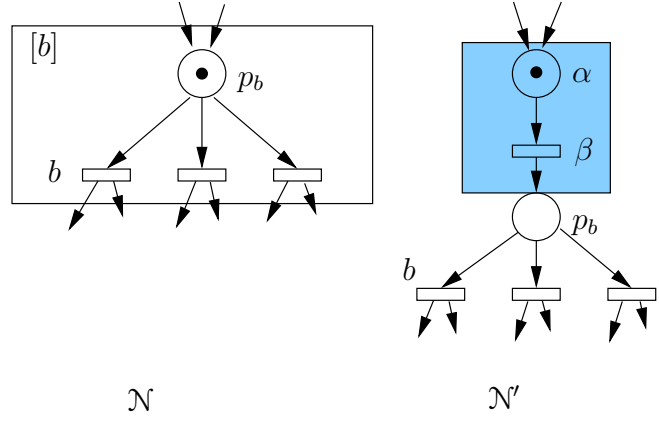}
  \caption{Introduction of a new, non-conflicting, transition.
  \label{newt}}
\end{center}
\end{figure} 

\begin{proof}
Here is a sketch of the proof.
If $b$ is non-conflicting, then the only transition in $[b]$ is $b$
and Theorem \ref{th:want} applies directly.

If $b$ is conflicting, then let $p_b$ be the only place in the cluster
$[b]$. We construct a new net  $\cN'$ by
introducing a new and non-conflicting transition $\beta$ and a place
$\alpha$ as shown in Figure \ref{newt}. 
If $M_0$ is the initial marking of $\cN$, we define the initial
marking $M_0'$ of $\cN'$ by 
\[
M_0'(p)=\begin{cases} M_0(p_b) & \mbox{ if } p=\alpha \\
                      0      & \mbox{ if } p=p_b \\
                      M_0(p) & \mbox{ otherwise }\:.
\end{cases}
\]

Now, $(\cN,M_0)$ and $(\cN',M_0')$ are equivalent in the following
sense. Let $\cP$ and $\cP'$ be the sets of places of $\cN$ and $\cN'$
respectively. Define the surjective mapping
\begin{eqnarray*}
\varphi : &\N^{\cP'} & \longrightarrow \ \ \N^{\cP} \\
          &  M'      & \longmapsto  \ \ M \:,
\end{eqnarray*}
with $M(p_b) = M'(\alpha) + M'(p_b)$ and $M(p)=M'(p)$ for $p\neq
p_b$. 
Clearly, if $M'$ is a reachable marking in  $\cN'$, then $\varphi(M')$
is a reachable marking in $\cN$.
 Furthermore, if $u$ is a firing
sequence leading to $M'$ in $\cN'$, then the word $v$
obtained from $u$ by removing all the instances of $\beta$ is a firing
sequence of $\cN$ leading to $\varphi(M')$.




Applying Theorem \ref{th:want} to $\cN'$ by blocking $\beta$ provides a
unique blocking marking $M'_\beta$. The marking $\varphi(M'_\beta)$ of
$\cN$ has all the required properties. 
\end{proof}

\begin{figure}[htbp]
  \begin{center}
\input{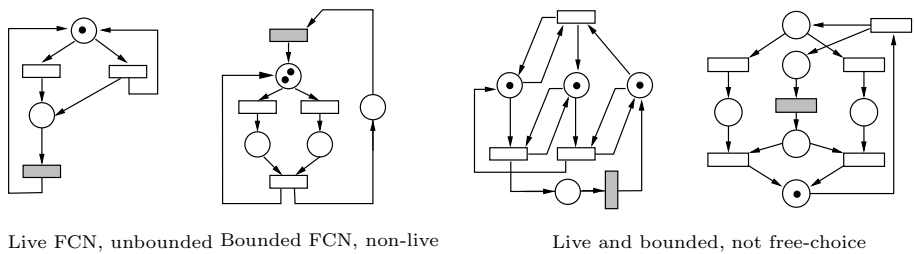}
    \caption{Several nets with non-unique blocking markings.
    \label{fi-ctrex}}
  \end{center}
\end{figure}

It is worth noting 
that none of the three assumptions in Theorem 
\ref{th:want} (liveness, boundedness, Free Choice property) can be dropped.
Figure \ref{fi-ctrex}  displays four nets which are
respectively non-live, unbounded and not Free Choice for the last two.
When blocking the transition in grey in these nets, several blocking
markings may be reached. More precisely, for each net in Figure
 \ref{fi-ctrex}, we have $|R_b(M_0)| \mgeq 2$ and $|R'_b(M_0)| \mgeq 2$. For the net on the left, 
we even have $|R_b(M_0)|=|R'_b(M_0)|=\infty$. 

Before we go on with the proof of Theorem  \ref{th:want}, we show
that the computation of the blocking marking is polynomial in the size
of the net.

\begin{proposition}\label{prop:compute}
Let $\cN$ be a bounded and live free-choice net and
let $b$ be a transition.
Then, computing the blocking marking $M_{[b]}$ is cubic in the size of $\cN$.
\end{proposition}

\begin{proof}
For each place $p$ not in $[b]$, choose a single output transition
$t(p)$ 
such that there exists a shortest path  from $p$ to $b$ that contains
$t(p)$. 
Note that such paths exist since $\cN$ is strongly connected.
Fire all transitions from 
$\trans_{[b]}:= \{t(p)\mid p\not\in [b]\}$,
in an arbitrary order
and as
often as possible; let $\sigma\in (\trans_{[b]})^*$ be such a firing 
sequence. Using the Pointing Allocation Lemma (\cite{DeEs}, Lemma 6.5),
 $\sigma$ is finite and leads to $M_{[b]}$.
By the Biased Sequence Lemma 
(\cite{DeEs}, Lemma 3.26), there exists another firing sequence $\tau \in
(\trans_{[b]})^*$ leading to $M_{[b]}$ whose length is at most $m T(T+1)/2$, where  $\cN$ is
$m$-bounded and  
$T$ is the number of its transitions. Now, according to Lemma
\ref{le-oneway} to be proved below,
we have $|\sigma|=|\tau|$.  

This yields a cubic time algorithm to find $M_{[b]}$.
The set of all shortest paths to a given node is found in
  quadratic time $O(T^2)$. 
Now, computing the marking reached after a firing 
sequence of length $O(m T^2)$ can be done in $O(m T^3)$ units of time.
\end{proof}

We now give the proof of Theorem \ref{th:want}. This proof is quite 
lengthy; since nothing that follows depends on this proof (of course, the 
{\em result} will be used frequently), readers are free to jump forward to 
Section \ref{se-fcerp}.

\begin{proof}[of Theorem \ref{th:want}]
Recall that $M_b$ is the {\em blocking marking} associated
with $b$.
It follows from the definition (see (\ref{eq-rb})) that we have
\begin{equation}\label{eq-basic}
\forall M \in R(M_0), \ \ R_b'(M) \subset R_b(M) \subset R_b(M_0)\:.
\end{equation}

According to Theorem \ref{th:TScover}, there exists a
covering of $\pn$ by T-components that we denote 
by $\tcomp_1,\ldots,\tcomp_n$.
The proof will proceed by induction on $n$.

\medskip

We assume first that $n=1$, that is, $\pn$ is a T-net. Note that all
the transitions 
are non-conflicting. 
The proof has four parts, each showing one of the following auxiliary
results. Given a  
transition $b$, one has for all $M\in R(M_0)$:
\[
1. \ \ R_b'(M)\neq \emptyset\:; \ \ \ 2. \ \ 
|R_b'(M)|=1\:; \ \ \ 3. \ R_b'(M)=R_b'(M_0)\:; \ \ \ 4. \  R_b(M)=R_b'(M) \:.
\]

\begin{enumerate}
\item The $T$-net $\pn$ is covered by circuits with a bounded number of
tokens, say $K$  
(Proposition \ref{DE318}). 
We block transition $b$ in the marking $M\in R(M_0)$. 
If $\gamma$  is a circuit of the covering containing $b$, it prevents any transition
in $\gamma$ from firing strictly more than $K$ times. 
Now, let $q$ be a transition such that there exist  circuits 
$\gamma_1,\ldots , \gamma_l$ from the covering such that $b$ belongs to $\gamma_1$,
$q$ belongs to $\gamma_l$, and $\gamma_i$ and $\gamma_{i+1}$ 
have a common transition for  $i=1,\ldots , l-1$. Then $q$ can fire at
most $l\cdot K$ times. Since $\pn$ is strongly connected, 
any transition can fire at most
$n\cdot K$ times, where $n$ is the number of circuits in the covering. 

\item The proof is almost the same  as for Lemma \ref{le-oneway}. 
Let us consider $M_1,M_2\in R_b'(M)$ with $M\move{\sigma_1}M_1$ and
$M\move{\sigma_2}M_2$ and $|\sigma_1|_b=  |\sigma_2|_b=0$.
We want to prove that $M_1=M_2$. 
There exist possibly several firing sequences with Parikh vectors
$\vec{\sigma}_1$ and $\vec{\sigma}_2$.  
Among these firing sequences, we choose the two with the longest
common prefix, and we denote them by 
$u_1=xv_1$ and $u_2=xv_2$ (recall that $\vec{u}_1=\vec{\sigma}_1$ and $\vec{u}_2=\vec{\sigma}_2$).
Let $\tilde{M}$ be such that  
$M\move{x} \tilde{M}$.
If $v_1=v_2=e$, then $M_1=M_2=\tilde{M}$. Assume that $v_1\neq e$ and
let $a$ be the first letter of $v_1$.  
Since
$|u_1|_a>0$, we deduce that $a\neq b$. 
The transition $a$ is enabled in $\tilde{M}$. Furthermore, by definition, 
$a$ is not enabled in $M_2$. This implies that the firing sequence $v_2$
must contain $a$; thus, we can set
$v_2=yaz$ with $|y|_a=0$. 
Since $a$ is 
enabled in $\tilde{M}$, it follows that $ayz$ 
is a firing sequence and $\tilde{M}\move{ayz} M_2$. 
To summarize, we have found two firing sequences $u_1$ and $u_2'=xayz$
with respective Parikh vectors 
$\vec{\sigma}_1$ and $\vec{\sigma}_2$ and with $xa$ as a common prefix.
\ This is  
a contradiction. 
\begin{figure}[htbp]
  \begin{center}
  \input{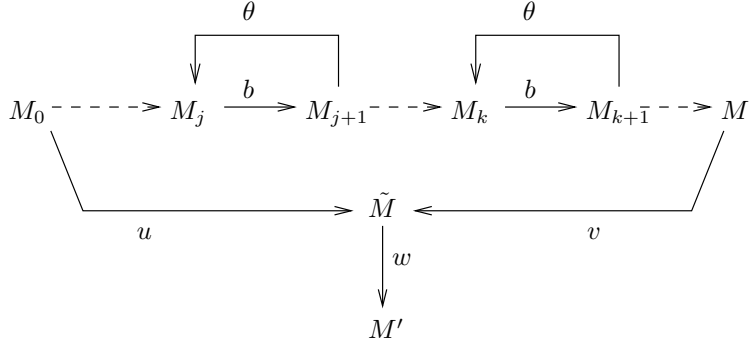}
    \caption{Using reverse firings to avoid $b$.\label{fig:negative}}
  \end{center}
\end{figure}

\item Let $\sigma$ be such that $M_0\move{\sigma}M$. If $|\sigma|_b=0$, it follows from
the previous point that 
$R_b'(M)=R_b'(M_0)$. Let us assume that $|\sigma|_b>0$. 
Let $\sigma=q_1\cdots q_n$ with $q_i\in \trans$ and $M_0 \move{q_1} M_1
\move{q_2}M_2 \cdots M_{n-1}\move{q_n} M_n=M$. 
Let $k$ be any index such that $q_k = b$, that is $M_{k-1}\move{b} M_k$. 
Using Propositions \ref{le:tinvglatt} and \ref{pr-new}, 
there exists a
firing sequence $\theta$ with Parikh vector $\vec{\theta} = (1,\dots,1) - \vec{b}$
and such that  $M_k \move{\theta} M_{k-1}$, that is $M_{k-1} \move{\theta^-} M_{k}$
(see Section \ref{ap:results}). 
By replacing every $b$  by $\theta^-$  in $\sigma$,
we get a generalized firing sequence 
$\sigma'\ \in \ ((\trans-\{b\})\cup(\trans^- -\{b^-\}))^*$ such
that $M_0\move{\sigma'}M$.

Using the rewriting rules in
\eref{eq-rewr} and applying Lemma \ref{le:lemmaA}, we find a generalized
firing sequence $\sigma''$ such that  
$\sigma'\rewr{*} \sigma''$ 
and such that $\sigma''=uv^-, u \in (\cT-\{b\})^*, v^- \in (\cT^- -\{b^-\})^*$. 
Let $\tilde{M}$ be the marking such that $M_0\move{u} \tilde{M}\mover{v} M$,
and $M'$ the unique element of $R_b'(\tilde{M})$. 
Since we have $M\move{v} \tilde{M}$ with $|v|_b=0$, we obtain that $R_b'(M)=\{M'\}$. 
By definition there exists a firing sequence $w\in (\cT-\{b\})^*$ such that 
$\tilde{M}\move{w} M'$. We deduce that we have $M_0\move{uw}M'$ with 
$uw\in(\cT -\{b\})^*$. This implies that  
$R_b'(M_0)=R_b'(M)$. 
The whole argument is illustrated in 
Figure \ref{fig:negative}.

\item Clearly we have $R_b'(M)\subset R_b(M)$. For the converse, consider 
$\tilde{M}\in R_b(M)$ and $u\in \cT^*$ such that
$M\move{u} \tilde{M}$. If $|u|_b=0$ then $\tilde{M}\in R_b'(M)$; so
assume $|u|_b>0$ and set 
$u=vbw$ with $|w|_b=0$. Let $\hat{M}$ be the marking such that 
$M\move{vb}\hat{M}$. By construction, we have
$\tilde{M}\in R_b'(\hat{M})$. Now, by point 3. above, this implies that 
$\tilde{M}\in R_b'(M)$. 
\end{enumerate}

Assume now that $\pn$ is covered by the $T$-components 
$\tcomp_1,\dots, \tcomp_n,$ with $n\mgeq 2$, and let $b$ be a non-conflicting transition. 
We also assume the  covering to be minimal, {\it i.e.} 
such that no $T$-component can be removed from it. Let $\cP_i$ and $\cT_i$ be the
places and transitions of $\tcomp_i$.  
Set  $\cN_+=\cN [\bigcup_{j=1}^{n-1} \cP_i\cup\cT_i]$ 
and $\cN_-=\cN [(\cP-\cP_+)\cup (\cT-\cT_+)]$, where $\cP_+$ and $\cT_+$ are
the places and transitions of $\cN_+$. 
Since
the covering is minimal,
the subnet $\cN_-$ is non-empty. 

Now, it is always possible to re-number the $\tcomp_i$'s  such
that $b\in \cN_+$ and $\cN_+$ is strongly connected.
This is shown  in the first part of the proof of Proposition 
7.11 in \cite{DeEs} (see also Proposition 4.5 in \cite{DeEs93}).


\medskip

On the other hand, the net $\cN_-$ has no reason to be connected. Let us denote by
$\kappa_1,\ldots ,\kappa_m,$ the  
connected components of $\cN_-$. According to Propositions 4.4. and 4.5
in \cite{DeEs93}, the nets $\kappa_j$ are CP-subnets of ${\pn}$ (see Section \ref{ap:results}). 
This result is also demonstrated in the second part of the proof of Proposition 7.11 in \cite{DeEs}. 
\begin{figure}
  \begin{center}
    \input{kappa.pstex_t}\label{fig:kappa}
    \caption{The net $\cN$ decomposed into $\pn_+$ and the CP-subnets $\kappa_1,\ldots,\kappa_m$.}
  \end{center}
\end{figure}

The decomposition of $\pn$ into $\pn_+$ and $\kappa_1,\ldots, \kappa_m,$ is
illustrated in Figure \ref{fig:kappa}.
By Proposition \ref{pr:single}, each  $\kappa_i$ has a single way-in
transition denoted $w_i$. Furthermore, $w_i$ has a unique input place that we denote $p_i$. 
Indeed, let us consider 
$p\in \ ^\bullet w_i$. We  have $p\in \cN_+$. Since $\cN_+$ is strongly connected, 
the set of successors of $p$  in $\cN_+$ is non-empty, and we conclude
that $|p^\bullet|>1$.
Now by the Free
Choice property, $p$ must be 
the only predecessor of $w_i$. 

\medskip

We first show that $R_b'(M_0)$ is non-empty. 
We proceed as follows. 

\begin{enumerate}
\item[a.] Using Proposition \ref{le:drumrum}, for all $i=1,\dots,m$, there exists 
a firing sequence $\sigma_{\kappa_i}\in (\cT_{\kappa_i} - \{w_i\})^*$ 
such that  no transitions in $\cT_{\kappa_i} - \{w_i\}$ is enabled after 
firing  $\sigma_{\kappa_i}$.
Let $M'_0$ be the marking obtained from $M_0$ after firing the sequence
$\sigma=\sigma_{\kappa_1}\cdots \sigma_{\kappa_m}$. No transition from $\cN_-$ is enabled in $M'_0$ except 
possibly the way-in transitions. 

\item[b.] Consider the subnet $(\cN_+, M_0'|_{\cN_+})$. 
We first prove that it is live and bounded. 
By 
Proposition \ref{le:drumrum}, under the marking $M'_0$, the net $\pn -
\kappa_m$ is a live and bounded 
Free Choice net. 
Now, we can prove that $\kappa_{m-1}$ is a CP-subnet of  $\pn - \kappa_m$ by
the same arguments 
as the ones used to prove that $\kappa_{m-1}$ is a CP-subnet of $\pn$. 
Again by  Proposition \ref{le:drumrum}, the net 
$\pn - (\kappa_m \cup \kappa_{m-1})$ is a live and bounded
Free Choice net.
By removing in the same way all the CP-subnets, we
finally conclude that $(\cN_+, M_0'|_{\cN_+})$
is a live and bounded
Free Choice net. 
Furthermore, $\cN_+$ admits 
a covering by $T$-components of cardinality $n-1$. 
By the induction hypothesis, there exists a firing sequence $x$ 
avoiding $b$ and which disables all the
transitions in  $\cT_+$ except $b$. 
Let $M_b$ be the marking of $\cN$ obtained from $M_0'$ after firing
$x$ (now viewed as a firing  
sequence of $\cN$). 

\item[c.] By construction, no transition from $\cT_+$ except $b$ 
is enabled in $(\cN,M_b)$. Let us prove that the transitions $w_i$ are also disabled in 
$M_b$. The transition $w_i$ is enabled if 
its input place $p_i$ is marked. Let $a$ be an output transition of $p_i$ 
belonging to $\cN_+$. 
By the free choice property, we have $\{p_i\}=\ ^\bullet{a} =\ ^\bullet{w_i}$. Since $a$ is conflicting
and $b$ is non-conflicting, we have 
$a\neq b$, which implies that $a$ is not enabled and that $p_i$ is not marked. 
\end{enumerate}

Clearly, the above proof also works for $(\cN,M)$ where $M\in R(M_0)$. Hence,
\begin{equation}\label{eq-basic2}
\forall M\in R(M_0), \ \ 
R_b'(M)\neq \emptyset\:.
\end{equation}
We have thus completed the first step  of the proof.
We now prove the following assertion. 

\medskip

Assertion $(A_0)$: {\em The $T$-net $\kappa_i$
has a unique {\em reference}
marking in which the only enabled transition is 
$w_i$. Furthermore, starting from the reference marking,
if $w_i$ is fired $h_i$ times, then  the other transitions
can fire  at most $h_i$ times. If all the transitions in $\kappa_i$ are fired
$h_i$ times, then the net goes back to the reference marking.}\\

{\bf Proof of $(A_0)$}:
First, according to Proposition 5.1 in \cite{DeEs93}, there is a reachable marking
$M_R$ where no transition  
is enabled except $w_i$. Now using the same argument as in point 2 above
(or as in the proof of Lemma \ref{le-oneway}), we obtain that $M_R$ is
the only such marking. According to Proposition 5.2 in \cite{DeEs93}, 
 $M_R$ satisfies: for all transition $q\neq w_i$, there is an unmarked 
path from $w_i$ to $q$.
The rest of assertion $(A_0)$ follows easily.\\


By assertion ($A_0$), the markings $M_0'$, $M_b$, and $M'_b$ coincide
on all the subnets 
$\kappa_i$.
We turn our attention to the following assertion.\\

Assertion $(A_1)$: {\em If $M'$ is a marking reachable from
$M'_0$ which coincides with $M'_0$ on all the places of  $\kappa_1,\cdots,
\kappa_m$,
then  the marking $M'$ is reachable from  $M'_0$ by
  firing and reverse firing of transitions from
 $\pn_+$  only.}\\

We first show how to complete the proof of the theorem,  assuming $(A_1)$. 
Consider $M'_b \in R_b(M_0)$. We want to show that $M'_b = M_b$. 
Apply $(A_1)$ to the marking $M'_b$: 
it is reachable from  $M'_0$ by
firing and reverse firing  of transitions from
 $\pn_+$ only. 
We have seen above that $(\pn_+, M_0'|_{\cN_+})$ is a live and bounded
Free Choice net. 
It follows readily that $(\pn_+, M_b'|_{\cN_+})$ is also live and bounded. 
Since $\pn_+$ admits a covering by $T$-components of cardinality
$n-1$, we can apply the induction 
hypothesis to $\pn_+$: if $M$ and $M'$ are two markings of $\pn_+$ such that
$M \move{q} M'$ or $M \move{q^-} M'$ for some $q$ in $\cT_+$,
then the blocking  markings reached from $M$ and $M'$  are the same.
By repeating the argument for all transitions (which are fired or
reverse fired) 
on the path from $M'_0|_{\cN_+}$ to $M'_b|_{\cN_+}$,
we get  that $M_b'|_{\cN_+} = M_b|_{\cN_+}$. It follows that $M_b' = M_b$, i.e.
$R_b(M_0)=\{M_b\}$. Coupled with the results in \eref{eq-basic} and \eref{eq-basic2},
it implies that $R_b(M)=R'_b(M)=\{M_b\}$ for any reachable 
marking $M$. The only remaining point consists in proving assertion $(A_1)$. 

\medskip

{\bf Proof of $(A_1)$}: 
Let $\tau$ be a firing sequence leading from $M_0'$ to $M'$ and
let $h_i=|\tau|_{w_i}$ for $i=1,\dots,m$.
The proof proceeds by induction on
$h=h_1+ \cdots + h_m$.
The case $h = 0$ is trivial, since, under $M'_0$, no transition in $\kappa_1,\ldots,\kappa_m,$
can fire without firing the way-in transitions first.

Now let us consider the case where $h_1+ \ldots + h_m > 0$.
Since $M'_0$ and $M'$ coincide on  $\kappa_1,\ldots,\kappa_m$,
it follows from $(A_0)$ that all the transitions in $\kappa_i$ have fired $h_i$ times
in the sequence  $\tau$.

Without loss of generality (by re-numbering the $\kappa_i$'s) we can assume
that the last way-in transition 
fired in the sequence $\tau$ is $w_1$.
By commuting the last occurrence of $w_1$ with the transitions
in $\tau$ which can fire independently of it, we can assume that all the
transitions
in  $\kappa_i$ for  $i=2,\cdots,m,$ have fired  $h_i$ times and  all the
transitions
in  $\kappa_1$  have fired  $h_1-1$ times before $w_1$ is fired for the
last time. This means that the 
 marking $M_1$ reached just before $w_1$ is
fired for the last time coincides with $M'_0$ on all the  $\kappa_i's$.

\medskip

Let $\tau_{\kappa_i}$ be a firing sequence of $\kappa_i$ 
leading from the reference marking of $\kappa_i$ to itself 
(see $(A_0)$). We have $|\tau_{\kappa_i}|_t=1$ for $t\in \kappa_i,$ and 
$|\tau_{\kappa_i}|_t=0$ otherwise (see $(A_0)$).
By further commutation of transitions which can fire independently, 
the sequence $\tau$ can be rearranged and decomposed as displayed in
(\ref{eq:reach1}), where arrows $\moved{}$  mean ``{\it only transitions in
$\kappa_1,\ldots,\kappa_m$ are fired}'';  arrows $\moveu{}$ mean 
``{\it only transitions in 
$\pn_+$ are fired}'';
 and arrows $\moven{} $ mean ``{\it only transitions and reverse
   transitions from  
$\cN_+$ are fired}'':
\begin{equation}
  \label{eq:reach1}
  M_0 \moved{\sigma} M'_0 
      \moven{v} M_1
      \moved{\tau_{\kappa_1}} M_2 
      \moveu{u} M'.
\end{equation}

The firing sequence  $M'_0\moven{v}M_1$, with $v$ being
a generalized firing sequence containing only (reverse) transitions
from $\cN_+$,  exists by the induction  
hypothesis on $(A_1)$.
In the subnet $\kappa_1$, the firing sequence $\tau_{\kappa_1}$ 
leads from the reference marking to itself. 
However, the sequence has some side effects in the net $\cN_+$, since a
token has been 
removed from the place $p_1$, and one token has been added in each output place 
of a way-out transition of $\kappa_1$. 
The challenge is now to ``erase'' this change in $\pn_+$ while  using
only transitions from $\pn_+$.

\medskip

To do this, consider the subnet ${\cal G} = \pn_+ \cup \kappa_1$.
We have proved in point b. above that the net $(\cN_+,M_0'|_{\cN_+})$ is a 
live and bounded Free Choice net. It follows clearly that ${\cal G}$ 
is live and bounded under the 
marking $M_0'|_{\cG}$. This implies that ${\cal G}$ is also live and
bounded under the  
marking $M_1|_{\cG}$ (since, in $\cN$, the marking 
$M_1$ is obtained from $M'_0$ by firing and reverse firing of transitions 
from $\cG$). 
By Theorem \ref{th:TScover}, the net $(\cG,M_1|_{\cG})$  can be
covered by $T$-components.  
Let $\cZ$ be a $T$-component of the covering which contains $w_1$.
By definition, $\cZ$ must also contain all the places in  $w_1^\bullet$.
Since $\cZ$ is strongly connected, it must contain the unique output transition
of each place in  $w_1^\bullet$. By repeating the argument, we get that
the  whole subnet $\kappa_1$  is included in $\cZ$.

\medskip

In the following, we play with the three nets $\cN$, $\cG$ and $\cZ$ (with
$\cZ\subset \cG\subset \cN$). 
To avoid very heavy notations, we use the same symbol for the marking in one
of the three nets and its restrictions/expansions to the other
two. For instance 
we use $M_1$ for $M_1,M_1|_{\cG}$ or $M_1|_{\cZ}$. 
We hope this is done without ambiguity. 

\medskip

Applying Theorem \ref{th-acti} to $(\cG,M_1)$, there exists a marking 
$M_3$ and a firing sequence $x$ such that $M_1\move{x} M_3$, 
the subnet $(\cZ,M_3)$ is live and $x$ contains no transition from $[\cZ]$. 
Recall that $[\cZ]$ is the cluster of $\cZ$. 
By construction, $x$ contains only transitions from $\cN_+$. 
In particular, the markings $M_1$ and $M_3$ coincide on the
subnet $\kappa_1$; moreover, no transition of $\kappa_1$ except possibly $w_1$
is enabled in $M_3$.  
Now we claim that $w_1$ is enabled in $M_3$. By definition of a cluster, 
the input place $p$ of $w_1$ belongs to $[\cZ]$, as well as all the
output transitions of $p$.  
We deduce that $x$ does not contain the output transitions of $p$, and
$w_1$ is enabled in $M_3$ 
since it was enabled in $M_1$. 

\medskip

Consequently, the sequence $\tau_{\kappa_1}$ is a firing sequence in $(\cZ,M_3)$.
Let $M_4$ be the marking defined by $M_3\move{\tau_{\kappa_1}}M_4$. 
Let $\cT_{\cZ}$ be the set of places of $\cZ$. We consider the vector 
$X\in \N^{\cT_{\cZ}}$
defined by $X_t=0$ if $t$ belongs to $\kappa_1$ and $X_t=1$ otherwise.
By construction and Assertion $(A0)$, we have $X + \vec{\tau}_{\kappa_1} = (1,\dots ,1)$. 
According to Proposition \ref{le:tinvglatt}, this implies that $M_4+ N_{\cZ}\cdot X = M_3$,
where $N_{\cZ}$ is the incidence matrix of $\cZ$. 
According to Proposition \ref{pr-new}, there exists a firing sequence $\theta$ 
of $(\cZ,M_4)$ such that $\vec{\theta}=X$. This implies that $\theta^-$ is a 
generalized firing sequence leading from $M_3$ to $M_4$. 

\medskip

Now we want to prove that $x$ is a firing sequence of $(\cN,M_2)$. 
The firing of $\tau_1$ involves only places from $\cZ$ 
(the places from $\kappa_1$, the input place of the way-in transition, and the output 
places of the way-out transitions). This implies that $M_1$ and $M_2$ coincide on the places which 
do not belong to $[\cZ]$. Now $x$ contains only transitions outside of
$[\cZ]$, and if  
$t$ is a transition outside of $[\cZ]$ then the input places of $t$ do
not belong to $[\cZ]$ either. 
 \begin{figure}[htbp]
  \begin{center}
\input{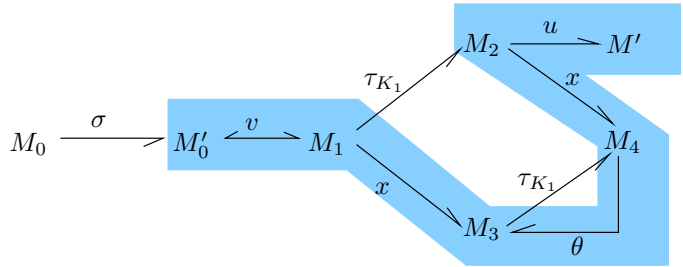}
    \caption{Proof of assertion $(A_1)$.\label{fi-summary}}
  \end{center}
\end{figure}
Since $x$ is a firing sequence of $(\cN,M_1)$, we deduce that it
is also a firing sequence of $(\cN,M_2)$. 
We have
\[
M_2 + N\cdot \vec{x}= M_1+ N\cdot (\vec{\tau}_{\kappa_1}+ \vec{x})=M_3 + N\cdot \vec{\tau}_{\kappa_1}=M_4\:.
\]
Hence we obtain $M_2\move{x}M_4$ and $M_4 \move{x^-} M_2$. Summarizing
the above steps, we have obtained  
that $\varpi=vx\theta^-x^-u$ is a generalized firing sequence leading from
$M'_0$ to $M'$ and involving only 
transitions and reverse transitions from $\cN_+$. 
This concludes the proof of $(A_1)$. The various steps are illustrated
in Figure \ref{fi-summary}, with the shaded area highlighting $\varpi$.

\end{proof}

\section{Blocking a Transition in a Routed FCN
}\label{se-fcerp}

In a live and bounded Free Choice net, only non-conflicting
transitions lead to a blocking marking, see Theorem \ref{th:want}.  
Furthermore, given any transition $b$ (even non-conflicting),
there exist in general infinite firing sequences not containing  $b$. 
This is for
instance the case in the net of Figure \ref{fi-example}. 
In this section, we introduce
routed Free Choice nets and we show that
there exists a blocking marking 
associated with {\em any} transition and that there is
no infinite firing sequence avoiding a given transition. 

\medskip

A {\em routed Petri net} is a pair $(\cN,u)$ where $\cN$ is a Petri
net (set  
of places $\cP$) and $u=(u_p)_{p\in \cP}$, $u_p$ being a function from $\N^*$ to
$p^\bullet$. 
For the places such that $|p^\bullet |\mleq 1$, the function $u_p$ is trivial. 
Below, it  will be convenient to consider $u_p$ as defined  either on
all the places or 
only on the places  
with several successors,
depending on the context. 
We call $u$ the {\em routing (function)}. To insist 
on the value of the initial marking $M$, we  denote the routed Petri
net by $(\cN,M,u)$.  

\medskip

A routed Petri net $(\cN,M,u)$ evolves as a Petri net except for the
definition of the 
{\em enabling} of transitions. A transition $t$ is {\em enabled} in
$(\cN,u)$ if it is enabled in $\cN$  
and if in each input place at least one of the tokens currently present
is {\em assigned}  to $t$ by $u$. 
The assignment is defined as follows: (1) in the initial marking of place
$p$, the number of tokens assigned  
to transition $t\in p^\bullet$ is equal to $\sum_{i=1}^{M_p} {\bf 1}_{\{u_p(i)=t\}}$
(where ${\bf 1}_A$ is the indicator function of $A$); 
(2) the $n$-th token to enter place $p$ during an evolution of the net is
assigned to transition $u_p(n+M_p)$, where the numbering of tokens entering $p$ is done 
according to the ``logical time'' induced by the firing sequence.

\medskip

Modulo the new definition of enabling of a transition, 
the definitions of {\em firing}, {\em firing sequence}, {\em reachable
  marking}, {\em liveness}, 
{\em boundedness} and {\em blocking transition} remain unchanged. 
We also say that a firing or a firing sequence of $\cN$ 
is  {\em compatible with} $u$ if it is also a firing or a firing sequence of 
$(\cN,u)$. 
Let $(\cN,M,u)$ be a routed 
Petri net and let us consider $M\move{\sigma}M'$; the resulting routed Petri net 
is $(\cN,M',u')$ where the routing $u'$ is defined as follows. In the
marking  
$M'$, the number of tokens of place $p$ assigned to transition $t\in
p^\bullet$ is equal to  
\begin{equation}\label{eq-assi}
\sum_{i=1}^{M'_p} {\bf 1}_{\{u_p'(i)=t\}} = \sum_{i=1}^{K} {\bf 1}_{\{u_p(i)=t\}} - |\sigma|_t, \ \ 
K=M_p+\sum_{t\in ^\bullet p} |\sigma|_t\:;
\end{equation}
and the $n$-th token to enter place $p$ is assigned to 
$u'_p(n+M_p')=u_p(n+M_p+\sum_{t\in ^\bullet p} |\sigma|_t)$. 
For simplicity and with some abuse, we use the notation
$(\cN,M',u)$ instead of $(\cN,M',u')$. We keep or adapt the notations
of Section \ref{se-prel}.  
For instance, the reachable markings of $(\cN,M',u)$ are denoted by
$R(M',u)$ (or $R(\cN,M',u)$).  
We also use the notations $R_b(M,u)$ and $R_b'(M,u)$ for the
analogs of the quantities defined in 
\eref{eq-rb}. 
For details on the semantics of routed Petri nets, see \cite{GaHa}. 

\medskip

\begin{figure}[htbp]
  \begin{center}
   \input{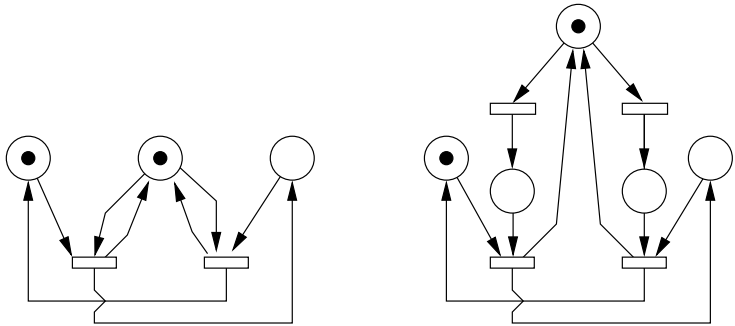}
    \caption{Compare the liveness of the routed and unrouted versions
      of the above Petri nets. \label{fi-ld}}
  \end{center}
\end{figure}

Clearly, we have $R(\cN,M,u)\subset R(\cN,M)$; hence, if $\cN$ is
bounded, so is $(\cN,u)$.
The converse is obviously false. 
The liveness of $\cN$ or $(\cN,u)$ does not imply the liveness of the other.
For instance, the Petri net on the left of 
Figure \ref{fi-ld} is live but its routed
version is live only for the routing $ababa\cdots$ ($a$ being the
transition on the left and $b$  
the one on the right). For the Petri net on the right of the same
figure, the routed version is live 
for the routing $ababa\cdots$ but the (unrouted) net is not live. 

We need an additional definition: the routing $u$ is {\em equitable} if 
\begin{equation}\label{eq-equitable}
\forall p\in \cP, \forall t \in p^\bullet, \ \ \sum_{i\in \N^*} {\bf 1}_{\{u_p(i)=t\}} = \infty\:.
\end{equation}
In words, a place that receives  an infinite number of tokens
assigns an infinite number of them  to each of
its output transitions.  
The next two results establish the relation between  the unrouted 
and routed behaviors of a net.

\begin{lemma}\label{le-bound}
Let $\pn$ be a Petri net.
The following statements  
are equivalent: \\
\hspace*{0.5cm} 1. $(\pn,u)$ is bounded for any routing $u$;  \\
\hspace*{0.5cm} 2. $\cN$ is bounded. 
\end{lemma}

\begin{proof}
Clearly, 2. implies 1. 
Assume that $(\cN,M_0)$ is unbounded. 
Classically, this implies that there exists 
$M_1\in R(M_0)$ and $M_2\in R(M_1)$ such that $M_2>M_1$. This is proved using a construction 
by Karp and Miller, see \cite{KaMi69} or Chapter 4 in \cite{reut}. 
Consequently, there exists a sequence of reachable markings $(M_i)_{i\in \N^*}$ 
and a firing sequence $\sigma$ such that $M_i\move{\sigma}M_{i+1}$ 
and such that the total number of tokens of $M_i$ is strictly increasing. 
Let $\sigma_0$ be such that $M_0\move{\sigma_0}M_{1}$ and let $\tau$ be the infinite sequence
defined by $\tau=\sigma_0\sigma\sigma\cdots$. 
Choose  
a posteriori a routing  $u$ compatible with
$\tau$. Clearly, $(\cN,u)$ is unbounded and we have proved that non-2. implies 
non-1.
\end{proof}

\begin{lemma}\label{le-viv}
Let $\pn$ be a
Free Choice net. The following
propositions  
are equivalent: \\
\hspace*{0.5cm} 1. $(\pn,u)$ is live for any 
equitable routing $u$; \\
\hspace*{0.5cm} 2. $\cN$ is live. 
\end{lemma}


\begin{proof}
First note that if $(\pn,u)$ is live then clearly $u$ must be equitable. 
Let us prove that 1. implies 2.
Let $M_0$ be the initial marking and consider $M \in R(\cN,M_0)$ and
an arbitrary  transition $q$ of
$\cN$.
Clearly there exists en equitable routing $u$ such that 
$M \in R(\cN,M_0,u)$. Since $(\cN,M_0,u)$ is live, $(\cN,M,u)$ is also live and 
there is a firing sequence of $(\cN,M,u)$ which enables $q$. The same
sequence enables $q$  
in $(\cN,M)$. 

Now let us prove that 2. implies 1.
We assume that there exists an equitable routing $u$ such that
$(\cN,u)$ is not live.
There thus  exists a transition $q$ which is never enabled 
in $(\cN,u)$, after some firing sequence $\sigma$. Set $X=\{q\}$. 
By equitability of the routing $u$, this implies that 
${}^\bullet q$ contains a place $p$ which receives only a finite number of tokens 
after $\sigma$.
Then 
the transitions in ${}^\bullet p$ fire at most a finite number of times after $\sigma$. 
Set $X= X\cup \{ p\}\cup {}^\bullet p$. 
For each one of the new transitions in $X$, we use the argument first
applied to $q$ and  
 repeat the construction recursively. 
Since the net is finite, this construction terminates and 
we end up with a set of nodes $X$. 
The set $X\cap \cP$ is non-empty and a siphon (see Section \ref{ap:results}). 
By construction, there is a finite firing sequence leading to an empty
marking in the 
siphon $X\cap \cP$. We deduce that the siphon cannot contain an
initially marked trap, hence $\cN$ 
cannot be live by Commoner's Theorem \ref{th:commoner} (this is where
we need the Free Choice assumption).
\end{proof}

\begin{lemma}\label{le-le}
Let $\pn$ be a live and bounded Petri net and let $u$ be an equitable
routing. For any  
infinite firing sequence $\sigma$ of the routed net $(\pn,u)$ and for any
transition $t$, we have $|\sigma|_t =\infty$. 
\end{lemma}

\begin{proof}
We say  that a transition $q$ is {\em $\sigma$-live}
if $|\sigma|_q=\infty$ and
{\em $\sigma$-starved} otherwise.
We are going to prove that all transitions are $\sigma$-live. 
Obviously, since $\sigma$ 
is infinite, it is not possible for all transitions to be $\sigma$-starved. 
Assume there exists a transition 
$s$ which is $\sigma$-live and a transition $t$ which is $\sigma$-starved.
Since $\pn$ is strongly 
connected by Theorem \ref{th:strongco}, there are places $p_1,\ldots,p_n$
and transitions  
$q_1,\ldots ,q_{n-1}$ such that 
$s=q_0\to p_1\to q_1 \to \cdots \to q_{n-1} \to p_n \to q_n=t$. 
There  exists an index $i$ such that
$q_i$ is $\sigma$-live and $q_{i+1}$ is $\sigma$-starved.
Since  $u$ is equitable, an infinite number
of tokens going through $p_{i+1}$ are routed towards $q_{i+1}$. By
assumption, $q_{i+1}$  
consumes only finitely many of them under $\sigma$, which implies that the
marking of $p_{i+1}$ is unbounded. 
This is a contradiction. 
\end{proof}

Using the above lemma, 
we obtain for routed Free Choice nets a stronger version of
Theorem \ref{th:want}: all transitions (not just clusters !)
 yield a blocking marking, provided
the routing is equitable. 

\medskip

\begin{theorem}\label{th:blockrouted}
Let $(\pn,M_0)$ be a live and bounded Free Choice net.
For any transition $b$, there
exists a blocking marking $M_b$  
such that for every equitable routing $u$ and all $M\in R(M_0,u)$, we have 
$R_b(M,u)=R_b'(M,u)=\{M_b\}$. 
\end{theorem}


The proof is postponed to the end of Section \ref{sse-expa}, 
where a more general version of the result is given (in Theorem
\ref{th-blockrouted2}). More precisely, we prove the result for 
the class of Petri nets
whose Free Choice expansion is live and bounded.

Here, we now prove some additional results on routed Petri nets to be
used in Section \ref{sto}.  


\begin{lemma}\label{le-oneway}
Consider a live and bounded routed Free Choice net $(\cN,M_0,u)$.
Let $b$ be a 
transition and $M_b$ the associated blocking marking. 
For any $n\in \N$, there exists a firing sequence $\sigma$ of $(\cN,M_0,u)$
such that $|\sigma|_b=n$ and $M_0\move{\sigma}M_b$.
If $\sigma$ and $\sigma'$ are firing sequences of $(\cN,M_0,u)$
such that $|\sigma|_b=|\sigma'|_b, M_0\move{\sigma}M_b,$ and $M_0\move{\sigma'}M_b$,
then we have $\vec{\sigma}=\vec{\sigma}'$. If $\tau$ and $\sigma$ are firing sequences
such that $|\tau|_b\mleq |\sigma|_b$, and $M_0\move{\sigma}M_b$, then we have 
$\vec{\tau}\mleq \vec{\sigma}$. 
\end{lemma}
\begin{proof}
The existence of $\sigma$ such that $|\sigma|_b=n$ and $M_0\move{\sigma}M_b$ follows by induction
from Theorem \ref{th:blockrouted}. 

We give the proof of the 
remaining points in the case $\sigma \in (\cT-\{b\})^*$. The general
case can be argued in a similar way. 
The argument is basically the same as for Part 2. of the proof of Theorem 
\ref{th:want}. 
Let $u_1$ and $u_2$ be two firing sequences of $(\cN,M_0,u)$ 
such that $\vec{u}_1=\vec{\sigma}, \vec{u}_2=\vec{\sigma}'$, and 
with the longest possible common prefix. We set $u_1=xv_1$ and $u_2=xv_2$
where $x$ is the common prefix. 
If $v_1=v_2=e$, then obviously $\vec{\sigma}=\vec{\sigma}'$.
Assume that $v_1\neq e$, and let $a$ be the first letter of $v_1$. 
Let $\tilde{M}$ be such that  
$M_0\move{x} \tilde{M}$.
Since
$|u_1|_a>0$, we deduce that $a\neq b$. 
The transition $a$ is enabled in $\tilde{M}$. Furthermore, by definition, 
$a$ is not enabled in $M_b$. 
However, in a routed net, once a transition 
is enabled, the only way to disable it is by firing it.
This implies that the firing sequence $v_2$ must contain $a$; so, set
$v_2=yaz$ with $|y|_a=0$. 
Since $a$ is 
enabled in $\tilde{M}$, it follows that $ayz$ 
is a firing sequence and $\tilde{M}\move{ayz} M_b$. 
To summarize, we have found two firing sequences $u_1$ and $u_2'=xayz$
leading to $M_b$, 
with respective Parikh vectors
$\vec{\sigma}$ and $\vec{\sigma}'$ and with a common prefix at least equal to
$xa$. This is  
a contradiction. 

Now let us consider a firing sequence $\tau \in (\cT-\{b\})^*$ 
and let $M'$ be such that $M_0\move{\tau}M'$. By Theorem 
\ref{th:blockrouted}, there exists a firing sequence $\theta$ of $(\cN,M',u)$ such that 
$\theta \in (\cT-\{b\})^*$ and $M'\move{\theta} M_b$. Applying the first part of the proof, 
we get that $\vec{\tau}+\vec{\theta}=\vec{\sigma}$. 
\end{proof}

\begin{lemma}\label{le-same}
Let $(\cN,M_0,u)$ be a routed Petri net admitting a deadlock
$M_d$. Then $M_d$ is the unique deadlock  
of $(\cN,M_0,u)$. If $\sigma$ and $\sigma'$ are firing sequences of $(\cN,M_0,u)$
such that $M_0\move{\sigma}M_d, M_0\move{\sigma'}M_d$, then we have $\vec{\sigma}=\vec{\sigma}'$.
Furthermore if $\tau$ is a firing sequence 
of $(\cN,M_0,u)$, then $\vec{\tau}\mleq \vec{\sigma}$. 
\end{lemma}

\begin{proof}
The argument mimics the one of the second point in Lemma \ref{le-oneway}
(which does not require 
using Theorem \ref{th:blockrouted} and is valid for any routed Petri net). 
Assume first that there exist deadlocks $M_d^1$ and $M_d^2$ with 
$M_d^1\neq M_d^2$.  Let $M_0\move{\sigma_1}M_d^1$ and $M_0\move{\sigma_2}M_d^2$,
and assume $\sigma_1$ and $\sigma_2$ have been chosen, among all pairs
of firing sequences 
with this property,
 so that the length of the common prefix $\sigma$ of $\sigma_1$ and $\sigma_2$
is maximal. Let  $M_\sigma$ be such that $M_0\move{\sigma}M_\sigma$. Then 
$M_\sigma\not\in\{M_d^1,M_d^2\}$. Let $q_1$ be the
transition following the prefix $\sigma$ on $\sigma_1$.
The tokens in $M_\sigma$ used by $q_1$ 
can not be used by any other transition since their routing will not be 
changed; hence those tokens remain untouched by the 
suffix, after $\sigma$, of $\sigma_2$. As a consequence,
if $\sigma_2=vq_1w$, then $q_1vw$ is also a firing sequence starting
from $M_\sigma$, which contradicts that $\sigma_1$ and $\sigma_2$ have
been choosen with the maximal common prefix. So,
we have $M_d^2\move{q_1}$, which contradicts that $M_d^2$ is a
deadlock. 

Now, let $M_0\move{\sigma}M_d, M_0\move{\sigma'}M'$ with 
$|\sigma |_q< |\sigma'|_q$ for some transition $q$. Choose $\sigma$, 
$\sigma'$, and $q$ with the above properties and  such that 
the common prefix $\bar{\sigma}$ of $\sigma$ and $\sigma'$ is of 
maximal length. 
Set $\sigma=\bar{\sigma}w$ and $\sigma'=\bar{\sigma}qw'$.
Clearly we have $|w|_q=0$. 
The same reasoning as above leads to conclude that $M_d\move{q}$, contradicting
the deadlock property. Therefore, we have
$\vec{\sigma'}\mleq\vec{\sigma}$. 
In the particular case $M'=M_d$, it follows that 
$\vec{\sigma}=\vec{\sigma}'$.
\end{proof}

\section{Stationarity in Stochastic Routed FCNs
}
\label{sto}

\subsection{Stochastic routed Petri nets}

A {\em timed routed Petri net} is a routed Petri net with {\em firing
  times} associated with transitions. 
(Here we do not consider {\em holding times} associated with places
for simplicity.  
As usual, this restriction is done without loss of generality. Indeed, 
a timed Petri net with firing and holding times can be transformed into an
equivalent expanded Petri net with only firing times.)
The firing semantics is defined as follows. The timed evolution of the marking
starts at instant 0 in the initial marking. 
Let $a$ be a transition with firing time $\sigma_a\in \R_+$, and which
becomes enabled at instant $t$.  
Then,
\begin{enumerate}
\item at instant $t$, the firing of $a$ {\em begins}: one token is
  {\em frozen} in 
each of the input places of $a$. A frozen token can not get involved in any
other enabling or firing;
\item at instant $t+\sigma_a$, the firing of $a$ {\em ends}: the frozen
  tokens are removed and  
one token is added in each of the output places of $a$.
\end{enumerate}

Obviously, this semantics makes sense only if a given token can not
enable several  
transitions simultaneously. In a routed Petri net, this is the case. With this
semantics, an enabled transition immediately starts its firing; we say
that the evolution is 
{\em as soon as possible}. Timed routed Petri nets were first
studied in \cite{BaCG}.  

\medskip

The firing times at a given transition may not be the same from firing
to firing.  
In general, 
the firing times at transition $a$ are given by a function 
$\sigma_a: \N^* \to \R_+$, 
the real number $\sigma_a(n)$ being the firing time for the $n$-th firing 
at transition $a$. 
The numbering of the firings is done according to the instant of
initiation of the firing (the ``physical time'').  
Let $u$ be the routing; recall that $u_p(n)$ is the transition 
to which $u$ assigns the $n$-th token to enter place $p$. 
Here again, we assume that the numbering of the tokens entering place
$p$ is done according 
to the ``physical time'' (as opposed to the untimed case, where the
numbering was done according  
to the ``logical time'' induced by the underlying firing sequence). 

\medskip

Let $(\Omega,\cS,P)$ be a probability space. From now on, all random
variables are defined  
with respect to this space. 
A {\em stochastic routed Petri net} is a timed routed Petri net where 
the routings and the firing times are random variables; more
precisely,  a quadruple $(\cN,M,u,\sigma)$ where $(\cN,M)$ is a Petri
net 
(places $\cP$ and transitions $\cT$), where $u=[(u_p(n))_{n\in \N^*},
p\in \cP]$ are the routing sequences, 
and where $\sigma=[(\sigma_a(n))_{n\in \N^*}, a\in \cT]$ are the
firing time sequences.  
Furthermore, we assume that
\begin{itemize}
\item for each place $p$, $(u_p(n))_{n\in \N^*}$ is a sequence of i.i.d. 
r.v. (the so-called {\em Bernoulli routing});
\item for each transition $a$, $(\sigma_a(n))_{n\in \N^*}$ is
a sequence of i.i.d. r.v. and $E(\sigma_a(1))< \infty$; 
\item the sequences $(u_p(n))_{n\in \N^*}$ and $(\sigma_a(n))_{n\in \N^*}$
  are mutually independent.  
\end{itemize}

For details and other approaches concerning stochastic Petri nets, see
for instance 
\cite{ABDFC,BauKri:96}.

\medskip

By the Borel-Cantelli Lemma, we have for any place $p$ and any transition
$t\in p^\bullet$:
\begin{eqnarray*}
P\left\{ \ \sum_{i=1}^{+\infty} {\bf 1}_{\{u_p(i)=t\}}=+\infty \ \right\} & =&
\begin{cases} 1 & \mbox{if } P\{u_p(1)=t\} >0 \\
              0 & \mbox{otherwise.}
\end{cases}
\end{eqnarray*}
When $\forall p\in \cP, \forall t\in p^\bullet, \  P\{u_p(1)=t\} >0$,
the random routing is said to be {\em equitable} (since it is
equitable in the sense of \eref{eq-equitable} 
for almost all $\omega\in \Omega$).  

\subsection{Existence of asymptotic throughputs}\label{sse-eat}

This section is devoted to the proof of the following result. 

\begin{theorem}
\label{th-order1}
Consider a live and bounded stochastic routed Free Choice net with 
an equitable routing. 
For any transition $b$, there exists a constant $\gamma_b\in \R_+$ such that 
\[ 
\lim_{n\to \infty} \frac{X_b(n)}{n} 
= \lim_{t\to \infty} \frac{t}{\cX_b(t)}= \gamma_b \ \ a.s. \mbox{ and in } L_1\:, 
\]
where $X_b(n), n\in \N^*,$ is the instant of completion of the $n$-th firing at transition $b$
and where $\cX_b(t), t\in \R_+,$ is the number of firings completed at transition $b$ up to time $t$. 
\end{theorem}

Generally and assuming existence, we define the {\em
  throughput} of a transition $b$ as the random variable $\lim_{t\to
  \infty} \cX_b(t)/t$ (the average number of firings per time unit). 
Theorem \ref{th-order1} states that the throughput of any transition
  exists and is almost surely a constant.


\medskip

To prove Theorem \ref{th-order1}, we need some preparations.
Let $\eN=(\cN, M,u,\sigma)$ with $\cN=(\cP,\cT,\cF,M)$ 
be a {\em live and bounded stochastic routed Free Choice net with an
  equitable random routing}  
(SRFC in the following). 
We select a transition $b$ and we denote by $M_b$ the associated
blocking marking.

\begin{lemma}  
\label{positive} 
Assume that $\sigma_b(n)=+\infty$ for $n\in \N^*$, the other firing times and the routings
being unchanged.     
Let $\tau$ be the first instant of the evolution when the marking
reaches $M_b$ ($\tau=\infty$ if 
 $M_b$ is never attained). The r.v. $\tau$ is a.s. finite 
and integrable.  
\end{lemma} 

\begin{proof} 
According to Theorem \ref{th:blockrouted}, we have $R'_b(\cN,M,u)=\{M_b\}$ which means 
precisely that there exists a firing sequence $x$ such that $|x|_b=0$ and 
$M\stackrel{x}{\longrightarrow} M_b$. Define
\[
T = \sum_{a\in \cT-\{b\}} \sum_{i=1}^{|x|_a} \sigma_a(i)\:.
\]
Let us consider the timed evolution of the Petri net and let $v$ be
the firing sequence up to 
a given instant $t\in \R_+$. Since $\sigma_b(n)=+\infty$, we have $|v|_b=0$. 
According to Lemma 
\ref{le-oneway}, this implies that $\vec{v} \mleq \vec{x}$. 
Due to the as soon as possible 
firing semantics, $\eN$ is non-idling: 
at all instant at least one transition is firing. 
Furthermore, if the marking is different from $M_b$, there is always
at least one transition 
other than $b$ which is firing. We deduce that if $t\mgeq T$, then  we must 
have $\vec{v}=\vec{x}$; 
in other words, we have $\tau \mleq T$. This 
shows in particular that $\tau$ is a.s. finite. 

To prove that  $\tau$ is integrable, we need a further argument. 
A consequence of Lemma \ref{le-oneway} is that $\vec{x}$ depends only
on the routings and not on 
the timings in the SRFC. This implies in particular that the
r.v. $\vec{x}$ is independent of  
the random sequences $(\sigma_a(n))_{n},a\in \cT$, and hence 
\begin{eqnarray}\label{eq:espformule} 
E(T)&=& \sum_{a\in \cT-\{b\}} E(|x|_a)E(\sigma_a(1)). 
\end{eqnarray}
We specialize the SRFC
to the case where all the firing times are exponentially distributed with
 parameter 1, i.e. 
$P\{\sigma_a(1)>z\}=\exp(-z)$. 
Let $M_t$ be the marking at instant $t$.
The process $(M_t)_t$ is a continuous time Markov chain with state space 
$R(M)$. Let $T_n$ be the instants of jumps of $M_t$ and set
$M_n=M_{T_n}$. Then $(M_n)_n$ is 
a discrete time Markov chain and $\sum_a |x|_a$ is precisely the time
needed by the chain  
to reach the marking $M_b$ 
starting from $M$. 
Using elementary Markov chain theory, we get that $E(\sum_a |x|_a)<\infty$.
Using (\ref{eq:espformule}), this yields the integrability of $\tau$.
\end{proof}

From now on, we assume without loss of generality that   
$M=M_b$, that is, the initial marking is the blocking marking.  
 Let $K$ be the enabling degree of $b$ in $M$: 
\begin{equation}
\label{eq-k}
K=\max \{ k \ : \ M \move{b^k} \}\:.
\end{equation}
By construction, we have $K\mgeq 1$.  
\begin{figure}[htbp]
  \begin{center}
\input{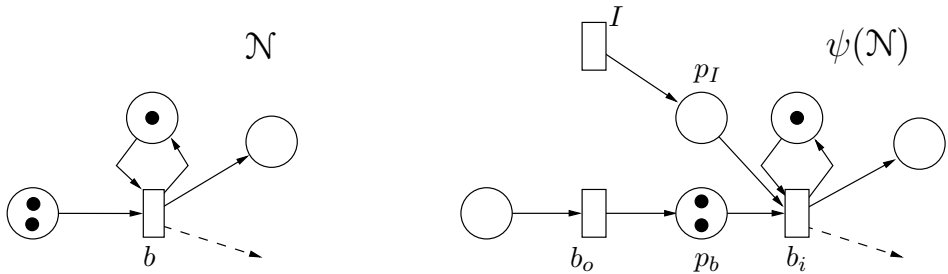} 
    \caption{Open Expansion of a Free Choice net.\label{fi-opex}}
  \end{center}
\end{figure}
We now introduce an auxiliary construction, the {\em Open Expansion} of 
an SRFC, which is characterized by an input transition $I$ without 
input places and a splitting of $b$ into an immediate transition
$b_o$ and a transition $b_i$ that inherits the firing duration of $b$.
\begin{definition}
The {\em Open Expansion} associated with $\eN$ and $b$ is the 
stochastic routed Free Choice net
$\psi(\eN)=(\psi(\cN), \psi(M), \psi(u), \psi(\sigma))$, where 
$\psi(\cN)$ is the net \\
$\psi(\cN)=(\psi(\cP),\psi(\cT),\psi(\cF),\psi(M))$, and
\begin{eqnarray*}
\bullet & \psi(\cP) &= \hspace*{4mm}\cP \ \cup \ \{p_b,p_I\} \\
\bullet & \psi(\cT) &= \hspace*{4mm}(\cT - \{b\} ) \ \cup \ \{I, b_{i}, b_{o}\}\\
\bullet & \psi(\cF) &
=\hspace*{4mm}(\cF -\{ (p,b)\in \cF, (b,p)\in\cF\})\\ 
&& \hspace*{8mm} \cup\ \{ (p,b_{o}) \ : \ (p,b)\in \cF, (b,p)\not\in \cF\}
\\ &&\hspace*{8mm} \cup
\{ (b_{i},p) \ : \ (b,p)\in \cF, (p,b)\not\in \cF\} \  \\
&&\hspace*{8mm}\cup \{ (b_i,p), (p,b_i) \ : (p,b) \in \cF, (b,p) \in \cF\} \\ 
&&\hspace*{8mm} \cup \
\{(I,p_I), (p_I, b_{i}), (b_{o},p_b), (p_b,b_{i})\}\\
\bullet & \psi(M)_p &= \hspace*{4mm}\left\{
\begin{array}{lcl}
 M_p &:&p \in \cP-({}^\bullet b) \\
M_p-K+K{\bf 1}_{\{p\in b^\bullet\}} &:& p \in ({}^\bullet b)\\
K&:& p=p_b\\
0&:& p=p_I
\end{array}
\right. \\
\bullet & \psi(\sigma)_a(n)&=\hspace*{4mm} \left\{
\begin{array}{lcl}
\sigma_a(n)&:& a \in (\cT - \{b\}) \\
\sigma_b(n) &:& a=b_i\\
0&:& a=b_o
\end{array} \right. \\
\bullet & \psi(u)_p(n)&= \hspace*{4mm} u_p(n) \:.\\
\end{eqnarray*}
\end{definition}
The construction is illustrated in Figure \ref{fi-opex}. Note that
$\psi(\cN)$ is neither 
live nor bounded. The marking $\psi(M)$ is a 
deadlock for the Petri net $\psi(\cN)$ (no transition is enabled). 

In the definition of $\psi(\eN)$,
we have not specified the value of $(\sigma_I(n))_{n}$. This is on purpose. 
Assume first that transition $I$ fires an infinite number of times at
instant 0  
($\forall n, \sigma_I(n)=0$). Then this {\em saturated} version of the net $\psi(\eN)$ behaves
exactly as $\eN$ (the firing times of $t\in \cT-\{b\}$ are the same in
the two nets and the 
firing times of $b_i$ in $\psi(\eN)$ are equal to the firing times of
$b$ in $\eN$).  
We are going to use this remark below. 

Assume now that $I$ fires a finite number of times at positive
instants. Then we can view  
$\psi(\eN)$ as a mapping
of the instants of (completion of) firings of $I$ into the instants of
(completion of) firings 
of $b_{o}$. Let us make this point more precise.

\medskip

Let $\cB$ be the Borel-$\sigma$-field of $\R_+$. 
A {\em (positive finite) counting measure} is a measure $a$ on $(\R_+,\cB)$
such that $a(C)\in \N$ for 
all $C\in \cB$. For instance, $a([0,T])$ can be interpreted as the 
number of events of a certain type occurring between times $0$ and $T$;
this will be used below.
We denote by $\cM_f$ the set
of counting measures. 
Given a set $E$, we denote by $\cM_f(E)$ the set of all couples $(m,\xi)$ where 
$m\in \cM_f$ and $\xi = (\xi_1,\dots,\xi_k), \ \xi_i\in E, k=m(\R_+)$. The elements of
$\cM_f(E)$ are called {\em marked counting measures}. 

\medskip

Set $\psi(\eN)_{[1]}=\psi(\eN)$. 
Assume that transition $I$ fires 
only once. 
According to Lemma
\ref{le-oneway}, transition $b_{o}$ will also fire once, and according to Lemma
\ref{positive},
the net will end up in the marking $\psi(M)$ after an a.s. finite time $\tau$. 
We define the random vector 
\[
\xi_1=[(u_p(1), \dots ,u_p(k_p)), p\in \psi(\cP); \ 
(\sigma_a(1), \dots ,\sigma_a(n_a)), a\in \psi(\cT)-\{I\}]\:,
\]
where  $n_a$ is the number of firings of transition $a$ up to time
$\tau$, and $k_p$ is the number of tokens  
which have been routed at place $p$ up to time $\tau$. 
Let us set $\psi(u)_{[2]}= [(\psi(u)_p(k+k_p))_{k\in \N^*}, p\in \psi(\cP)]$ and
$\psi(\sigma)_{[2]}= [(\psi(\sigma)_a(n+n_a))_{n\in \N^*}, a\in \psi(\cT)-\{I\}]$.
Now, let $\psi(\eN)_{[2]}= (\psi(\cN),\psi(M), \psi(u)_{[2]}, \psi(\sigma)_{[2]})$,
still with the assumption that $I$ fires only once. 
We define the random  vector $\xi_2$ associated with $\psi(\eN)_{[2]}$ in
the same way as we defined  
the random vector $\xi_1$ associated with $\psi(\eN)_{[1]}$. 
By iterating the construction, we define
$(\xi_n)_{n\in \N^*}$. Obviously the sequence $(\xi_n)_{n\in \N^*}$ is i.i.d.

\medskip

Consider again the SRFC $\psi(\eN)$, now
with the assumption that transition $I$ fires a finite number of times, 
say $k$.
According to Lemma
\ref{le-oneway}, the transition $b_{o}$ will also fire $k$ times, and
according to Lemma 
\ref{positive},
the net will end up in the marking $\psi(M)$ after an a.s. finite time $\tau_k$. 
It follows from Lemma
\ref{le-oneway} that the set of firings and routings used up to time
$\tau_k$ is precisely the union of
the ones in $\xi_1,\dots , \xi_k$ (although the order in which they are
used may differ from the  
one induced by $\xi_1, \dots , \xi_k$). 
Assume furthermore that the instants where firings of $I$ start are
deterministic and given 
by a counting measure $a\in \cM_f$, and set $\xi=(\xi_1,\dots ,
\xi_k)$. Then $(a,\xi)$ belongs to  
$\cM_f(E)$ for an appropriate set $E$. Now let us set 
\begin{eqnarray*}\Phi:\cM_f(E)&\to& \cM_f(E)\\
(a,\xi)&\mapsto &(b,\xi),
\end{eqnarray*}
 where $b$ is
the counting measure of the instants of completions of the firings of $b_o$.

\medskip

We will now need some operations and relations on counting measures.
\begin{itemize}
\item For $a\in \cM_f$, set $|a|=a(\R_+)$, the number of points of the
counting measure. \item For 
$\alpha=(a,\mu)\in \cM_f(E)$, set $|\alpha|=|a|$. 
\item For $a\in \cM_f$, define the smallest point $\min(a)=\inf \{t \ : \ a(\{t\})\mgeq 1\}$ and
\item[$\quad$]the largest point $\max(a)=\sup\{t \ : \ a(\{t\})\mgeq 1\}$. \item For $\alpha=(a,\mu)\in \cM_f(E)$, set
$\max(\alpha)=\max(a)$ and $\min(\alpha)=\min(a)$. 
\item For $a,b\in \cM_f$, define $a+b\in \cM_f$ by $(a+b)(C)=a(C) + b(C)$. 
\item For $\alpha,\beta \in \cM_f(E), \alpha=(a,\mu),\beta=(b,\nu), \max(a) < \min(b)$, \item[$\quad$]
let $\alpha + \beta \in \cM_f(E)$ be given 
by $\alpha+\beta=(a+b, (\mu,\nu))$. 
\item For $a\in \cM_f, t \in \R_+$, 
define $a+t \in \cM_f$ by $(a+t)(C)=a(C-t)$, \item[$\quad$]and if
 $\alpha=(a,\xi)\in \cM_f(E),t \in \R_+$, 
set $\alpha+t= (a+t,\xi)$. 
\end{itemize}
Define a partial order on $\cM_f$ as follows. For $a,b\in \cM_f$, 
\[
a\mleq b \ \mbox{ if } \
\forall x\in \R_+, \ a([x,\infty)) \mleq b([x,\infty))\:.
\]
Similarly, define a partial order on $\cM_f(E)$ as follows: For $\alpha,\beta\in \cM_f(E)$ and $\alpha=(a,\mu), 
\beta=(b,\nu)$, let $\alpha \mleq  \beta$ if   $a\mleq b$ and $\mu$ is a ``suffix''
of $\nu$:
\[
\alpha \mleq  \beta \ \mbox{ if } \ a\mleq b \mbox{ and } \mu_{|a|}=\nu_{|b|}, \ \mu_{|a|-1}=\nu_{|b|-1}, \dots ,
\ \mu_{1}=\nu_{|b|-|a|+1}\:.
\]

The mapping $\Phi: \cM_f(E) \longrightarrow \cM_f(E)$ is {\em monotone-separable},
i.e.,  satisfies the following properties:
\begin{enumerate}
\item {\bf Causality}: $\alpha\in \cM_f(E)\ \implies\ |\Phi(\alpha)|=|\alpha|$ and $\Phi(\alpha) \mgeq \alpha$;
\item {\bf Homogeneity}: $\alpha\in \cM_f(E), x\in \R_+\ \implies \ \Phi(\alpha+x)=\Phi(\alpha)+x$;
\item {\bf Monotonicity}: $\alpha,\beta \in \cM_f(E), \ \alpha\mleq \beta \implies \Phi(\alpha) \mleq \Phi(\beta)$;
\item {\bf Separability}:\\
\hspace*{8mm} $\alpha,\beta \in \cM_f(E), \max(\Phi(\alpha)) \mleq \min(\beta) \implies 
\Phi(\alpha+\beta)=\Phi(\alpha) + \Phi(\beta)$.
\end{enumerate}

The monotone-separable framework has been introduced 
in \cite{BaFo93b}. 
Actually, the setting used here is the one
proposed in \cite{bona} and differs slightly from the one in \cite{BaFo93b}.
The above  properties of $\Phi$ are proved in a slightly different and 
more restrictive
setting in \cite{BaFG}, Section 5. 
However, the arguments remain essentially the same. Consequently, we 
provide only an outline of the proof of the  monotone-separable property of 
$\Phi$.

\medskip

The argument is based on the equations satisfied by the {\em daters}
associated with  
the net. For $a\in \psi(\cT), n\in \N^*$, let $X_a(n)$ be the $n$-th
instant of completion  
of a firing at transition $a$ with $X_a(n)=+\infty$ if 
$a$ fires strictly less than $n$ times. It is also 
convenient to set $X_a(n)=0$ for $n\mleq 0$.
The variables $X_a(n)$ are called the {\em daters} associated
with the SRFC. 

\medskip

Assume that $I$ fires $k$ times, the instants of firings being 
$0\mleq x_1\mleq \cdots \mleq x_k$. 
Given a transition $a$ and a place $p\in {}^\bullet a$, we define 
$\nu_{pa}(n)=\min\{k \ : \ \sum_{i=1}^k {\bf 1}_{\{u_p(i)=a\}}=n\}$. The daters satisfy the
following recursive equations, 
see \cite{BaCG} for a proof:
\begin{eqnarray*}
\forall\ n\ >\ k: \hspace*{20mm}
X_I(1)=x_1, \dots , X_I(k)=x_k, \ \ X_I(n)=\infty ; \hspace*{15mm} \\
 \forall\ a\ \in\ \psi(\cT)-\{I\}:  \hspace*{86.5mm}\nonumber\\
\hspace*{-10mm}
X_a(n)= \left\{ \max_{p\in {}^\bullet a} \left[ 
\min_{\footnotesize
(n_i,i\in {}^\bullet p) :
 \ M_p+ \sum_{i\in {}^\bullet p} n_i = \nu_{pa}(n)
} \ 
\max_{i\in {}^\bullet p} \ X_i(n_i)
\right] \right\} + \sigma_a(n)\:.\normalsize
\end{eqnarray*}
Playing with the above equations, it is not difficult (although
tedious) to prove that  
the operator $\Phi$ is monotone-separable. 

\medskip

Assume that $I$ fires exactly $k$ times with
all the firings occurring at instant 0. The corresponding 
marked counting measure is
$\alpha_k=((0,\dots ,0) \ ; \ (\xi_1,\dots ,\xi_k))$. 
Given that $\Phi$ is monotone-separable and that $(\xi_n)_{n\in \N^*}$ is i.i.d.,
we obtain using directly the results in \cite{BaFo93b,bona} 
that there exists $\gamma_b \in \R_+$ such that 
$\lim_n \max(\Phi(\alpha_n))/n = \gamma_b$ a.s. and in $L_1$. 

\medskip

We have seen above that the firings of $b_i$ in the saturated version of $\psi(\eN)$
coincide with the ones of $b$ in $\eN$. 
More precisely, consider $k>K$ (we recall that $K$ is defined in \eref{eq-k}) and let
$b_1\mleq \cdots \mleq (b_k=\max(\Phi(\alpha_k)))$ be the points of the counting measure 
of $\Phi(\alpha_k)$. 
The  net $\psi(\eN)$ with input $\alpha_k$ coincides with 
$\eN$ up to the instant $b_{k-K}$. Now it follows from Lemma \ref{positive}
that $E[b_k-b_{k-K}]< \infty$.
This implies in a straightforward way that
$\lim_k X_b(k)/k = \lim_k \max(\Phi(\alpha_k))/k = \gamma_b$ a.s. and in $L_1$. 
This concludes the proof of Theorem \ref{th-order1}.

%

\subsection{Computation of the asymptotic throughputs}

The section is devoted to proving 
that the limits $(\gamma_a,a\in \cT)$ in Theorem 
\ref{th-order1} can be explicitly computed up to a multiplicative
constant. 

\begin{proposition}\label{pr-comput}
The assumptions and notations are the ones of Section \ref{sse-eat}
and Theorem \ref{th-order1}.  
The constants $\lambda_a=\gamma_a^{-1}, a \in \cT,$ are the throughputs at the transitions. 
Let us define the matrix $R=(R_{ij})_{i,j\in \cT}$ as follows:
\[
R_{ij}= \begin{cases} \frac{1}{|{}^\bullet j|}  \sum_{p : i\to p \to j} P\{u_p(1)=j\}
& \mbox{if } \exists p\in \cP, i \to p \to j\:.\\
0 & \mbox{otherwise}\:.
\end{cases}
\]
The matrix $R$ is irreducible, its spectral radius is 1, and there is a 
unique vector $x=(x_a,a \in \cT), x_a\in \R_+^*, \sum_{a} x_a=1$, such that 
$xR=x$. The vector $(\lambda_a, a \in \cT)$ is proportional to $x$, i.e., 
there exists $c\in \R_+^*\cup\{\infty\}$ such that $\lambda_a = c x_a$ for all $a\in \cT$. 
\end{proposition}

\begin{proof}
 If there exists a transition $a$ such that
$\lambda_a=\infty$, then clearly  
$\lambda=(\lambda_a, a\in \cT)=(\infty,\dots ,\infty)$ 
since the net is bounded. 
We assume first that the constants $\lambda_a$ are finite (the constants
$\gamma_a$ are strictly positive).  

We recall that for a transition $a$, the {\em counter} $\cX_a(t)$ is
the number of firings completed 
at transition $a$ up to time $t$. We also define for all $a\in \cT$ and
$p\in {}^\bullet a$, the  
{\em counter} $\cY_{pa}(t)$ which counts the number of tokens assigned
by the place $p$ to the transition $a$  
up to time $t$. We have
\begin{equation}\label{eq-step1}
\cX_a(t) \mleq \cY_{pa}(t) \mleq \cX_a(t) + \overline{M}_p\:,
\end{equation}
where $\overline{M}_p$ is the maximal number of tokens in place $p$
(which is finite since 
the net is bounded). 
We also have
\begin{equation}\label{eq-step2}
\cY_{pa}(t) =  \sum_{i=1}^{K(t)} {\bf 1}_{\{u_p(i)=a\}}, \ \ K(t) = M_p+ \sum_{b\in {}^\bullet p} \cX_b(t)\:.
\end{equation}
Going to the limit in \eref{eq-step1} and \eref{eq-step2}, we get
\[
\lambda_a=\lim_t \frac{\cX_a(t)}{t} = \lim_t \frac{\cY_{pa}(t)}{t}= \lim_t
\frac{\sum_{i=1}^{K(t)} {\bf 1}_{\{u_p(i)=a\}}}{K(t)} 
\times \frac{K(t)}{t}\:.
\]
Applying Theorem \ref{th-order1} and the Strong Law of Large Numbers, we obtain
\[
\lambda_a= P\{u_p(1)=a\}  \sum_{b\in {}^{\bullet}p} \lambda_b\:.
\]
Since the above equality holds for any $p\in {}^{\bullet}a$, we deduce 
\[
\lambda_a= \frac{1}{|{}^{\bullet}a|} \sum_{p\in {}^{\bullet}a} P\{u_p(1)=a\}\sum_{b \in {}^{\bullet}p} 
\lambda_b\:.
\]
The above equality can be rewritten as 
$\lambda=\lambda R$, where $R$ is the matrix 
defined in the statement of the Proposition. 

Since the Petri net is strongly connected, it follows
straightforwardly that $R$ is irreducible.  
The Perron-Frobenius Theorem (see for instance \cite{BePl}) states
that $R$ has a unique (up to a multiple) eigenvector with coefficients
in $\R_+^*$, and that the associated eigenvalue is the spectral
radius. We conclude that the spectral radius of $R$ is 1, and that 
$\lambda$ is defined up to a multiple by the equality $\lambda =\lambda R$. 

\medskip

It remains to consider the case where $(\lambda_a,a\in \cT)=(\infty,\dots ,\infty)$. 
The only point to be proved is that
$R$ is of spectral radius 1. In this  case, the statement of the
Proposition holds with 
 constant $c=\infty$. 
However, the matrix $R$ depends only on the routing characteristics
and not on the firing times.  
Modify the stochastic routed net by setting  all the firing times
to be identically equal 
to 1. Then the new throughputs belong to $\R_+^*$. The first part of
the proof applies, the  
vector of throughputs is a left eigenvector associated with the
eigenvalue 1, and we conclude 
that the matrix $R$ is indeed of spectral radius 1. 
\end{proof}\hspace*{2cm} 

A consequence of Proposition \ref{pr-comput} is that the ratio $\lambda_a/\lambda_b, a,b\in \cT,$ 
depends only on the routings of the models and not on the timings. On
the other hand, the multiplicative constant  
$c$ of Proposition \ref{pr-comput} depends on the timings. A concrete
application of Proposition 
\ref{pr-comput} is proposed in Example \ref{ex-comput}.

\medskip

The vector  $\lambda = (\lambda_a,a\in \cT)$ is a strictly positive
and real-valued $T$-invariant of the net, 
that is, a solution of $N \lambda = 0$, where $N$ is the
incidence matrix of the net. 
The vector $\lambda$ is a particular  $T$-invariant, distinguished by 
its connection
with the routing probabilities.

\medskip
An interesting special case is the one of live and bounded stochastic
routed T-nets.  
For this restricted model, Theorem \ref{th-order1} was proved in \cite{bacc92} 
(see also \cite{BCOQ}) with the additional result
that $(\lambda_a, a\in \cT)=(\lambda,\dots , \lambda)$.
This is consistent with Proposition \ref{pr-comput}.
Indeed, for a T-net, the matrix $R$ is such that 
$(1,\dots ,1)= (1,\dots ,1)R$,
which implies according to Proposition \ref{pr-comput} that 
$(\lambda_a, a\in \cT)=(\lambda,\dots , \lambda)$. This is also
 consistent with Proposition \ref{le:tinvglatt}.

It is well known that the value of $\lambda$ is hard to compute or even to
approximate in $T$-nets, 
see \cite{BCOQ}, Chapter 8. 
We conclude that for a general SRFC the multiplicative constant 
$c$ of Proposition \ref{pr-comput} must be even harder to compute or
approximate. Note, however, that this constant can be computed
for a fluid approximation of the net, 
when the firing times are all deterministic, by using dynamic
programming and Howard-type algorithms, see \cite{CohGauQua}.

\subsection{Beyond the i.i.d. assumptions}\label{sse-notiid} 

The monotone-separable framework is designed to deal with more general
than i.i.d. 
stochastic assumptions. In our case, simply by using the results in
\cite{BaFo93b,bona},  
we obtain the same results as in Theorem \ref{th-order1} under the
following assumptions: 
the sequence $(\xi_n)_n$ is stationary and ergodic, and the r.v. $\tau$
defined in Lemma \ref{positive} 
is a.s. finite and integrable. Proposition \ref{pr-comput}  also
holds under the  
generalized assumptions. 

However, an even more general setting is to assume that $(\xi_n)_n$ is
stationary and ergodic, and that all the firing times are
integrable. The remaining task is then to prove that $\tau$ is
integrable. It is feasible for T-nets, unbounded Single-Input FCN and
Jackson networks (see \cite{bacc92,BCOQ,BaFG,BaFo94,BaFM96}). 
We should mention
that at least in the case of bounded Jackson networks proving
$E(\tau)<\infty$  is already quite intricate \cite{BaFM01}. 
For live and bounded Free Choice nets, we believe  that $\tau$ is
always integrable, but the proof is outside  the scope of this
paper.

\subsection{Stationary regime for the marking} 
\label{sse-marking}
 
The existence of asymptotic throughputs for all the transitions 
can be seen as a `first order' result. A more precise, `second order',
result would be the existence  
and uniqueness of
a stationary regime for the marking process; we discuss this type of
result here. 

\medskip

The model is the same as in Theorem \ref{th-order1} and $M_b$ is the
blocking marking 
associated with a transition $b$. 
We make the following additional assumptions: \\
(i) in the marking
$M_b$, the {\em enabling degree} of $b$ is equal  
to 1, i.e., $\min_{p\in {}^\bullet b} (M_b)_p=1$;\\
(ii) the distribution of  $\sigma_b$ is unbounded,
i.e., $P\{\sigma_b(1) > x\} >0,\quad \forall x \in \R_+$.\\

Consider the continuous time and continuous state space Markov process
$(X_t)_t$ formed by the  
marking and the residual firing times of the ongoing firings at instant $t$. 
Let $(T_n)_{n}$ be the instants when the
marking changes, and let $Y_n=X_{T_n^-}$. Then $(Y_n)_n$ is a Markov
chain in discrete time.  
Under the above assumptions, it is not difficult to prove that
$\{(M_b,0)\}$ is a regeneration 
point for $(Y_n)_n$. It follows using standard arguments that
$(Y_n)_n$ and $(X_t)_t$ have a unique stationary regime. 

\medskip

This result calls for some comments.
\begin{itemize}
\item 
Assumption (i) is always satisfied if transition $b$ is
recycled ({\it i.e.} $\{b^\bullet\}\cap {}^\bullet\{  b\} = \{ p_b\}$ where place
$p_b$ has an initial marking equal to $1$).
This is equivalent to the assumption that transition $b$ operates like a 
single server queue. 
\item
Closed Jackson networks are a subclass of live and bounded Free Choice
nets (in which assumption (i) is always satisfied). 
Cyclic networks are a subclass of closed Jackson networks. 
In \cite{boro86,sigm,KaMa92}, second order results 
for closed Jackson networks are proved. The proofs are basically the
same as the one sketched above.  
In the specific case of cyclic networks, the second order results hold
true under much weaker  
assumptions \cite{bamb,KaMa94,mair97}. This shows that conditions such
as (i) and (ii) are only sufficient 
conditions for the existence and uniqueness of stationary regimes. 
\item
When removing assumption (i), it becomes much more intricate to get 
second order results under reasonable sufficient conditions.
For instance, second order results can be obtained if the firing time of $b$ is
exponentially distributed. 
\end{itemize}

\section{Some Extensions}\label{se-fce}

\subsection{Extended Free Choice nets}

\begin{figure}[htbp]
  \begin{center}
\input{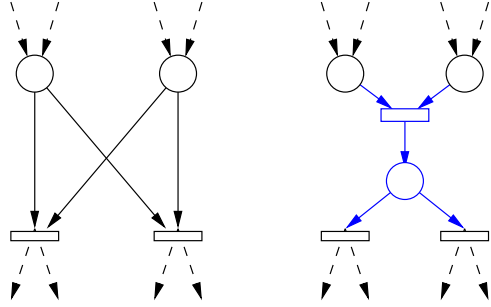}
    \caption{Transformation of an Extended Free Choice net into a Free
      Choice net.\label{fig:efcn}} 
  \end{center}
\end{figure}

It is common in the literature to consider {\em Extended Free Choice
  nets} (EFCN) 
defined as follows:
$\forall q_1,q_2\in \trans, \ 
p \in {}^\bullet q_1 \cap {}^\bullet q_2 \Rightarrow {}^\bullet q_1 =  {}^\bullet q_2$ 
(this is even the definition of {\em Free Choice nets} in \cite{DeEs}). 
The results in Theorem \ref{th:want} hold for EFCN. Indeed, given an
EFCN, one can apply  
Theorem \ref{th:want} to the Free Choice net obtained from the EFCN 
by applying the local transformation illustrated on Figure \ref{fig:efcn}.

On the other hand, the results from Sections \ref{se-fcerp} and
\ref{sto} do not apply to EFCN. 
In fact, the routed version of a live and bounded EFCN is in general not live. 

\subsection{Petri nets with a live and bounded Free Choice expansion}
\label{sse-expa}

\begin{figure}[htbp]
  \begin{center}
\input{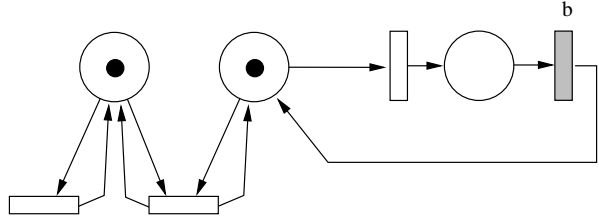}
    \caption{Petri net without a blocking marking.\label{fi-counter}} 
  \end{center}
\end{figure}

In this section, we consider the class of Petri nets having a live and bounded
{\em Free Choice expansion}. This class is strictly larger than the
one of live and  
bounded Free Choice nets (and strictly smaller than the one of live
and bounded Petri nets).  
The results related to routed nets
in Sections \ref{se-fcerp} and \ref{sto} extend to this class. 
On the other hand, Theorem \ref{th:want} can not be
extended to this class. As an illustration, the Petri net in Figure
\ref{fi-counter} has a live
and bounded Free Choice expansion and transition $b$ is non-conflicting,
but there exists no blocking marking associated with $b$. 

\begin{definition}
\label{def:phi}
Given a Petri net $\cN=(\cP,\cT,\cF,M)$, we define its {\em Free
  Choice expansion}  
$\transf(\cN)=(\transf(\cP),\transf(\cT),\transf(\cF),\transf(M))$ as follows:
\begin{itemize}
\item $\transf(\cP)=\cP \cup \{s_{pq} \ : \ p\in \cP, q \in p^\bullet\}$; 
\item $\transf(\cT)=\cT \cup \{t_{pq} \ : \ p\in \cP, q \in p^\bullet\}$; 
\item $\transf(\cF)= \cF \cup 
\{(p,t_{pq}), (t_{pq},s_{pq}),(s_{pq},q) \ : \ p\in \cP, q \in p^\bullet\}$;  
\item $\transf(M): \ \forall p\in \cP, \transf(M)_p=M_p, \ \forall p\not\in \cP, \transf(M)_p=0$. 
\end{itemize}
\end{definition}

\begin{figure}[htbp]
  \begin{center}
\input{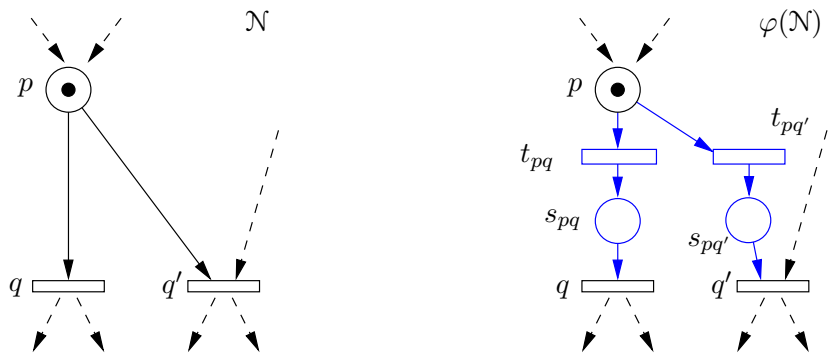}  
    \caption{Free Choice expansion of a Petri net.\label{fig:rfcn}}
  \end{center}
\end{figure}

Note that $\transf$ acts in a functional way (its components
mapping sets to sets), which justifies our
notation.
Obviously, the resulting net $\transf(\cN)$ is Free Choice. 
An example of this transformation is displayed 
in  Figure \ref{fig:rfcn}. 

It is easy to see that $\transf(\cN)$ is bounded if and only if $\cN$
is bounded.  
Liveness is more subtle. If $\transf(\cN)$ is live then clearly $\cN$
is also live.  
On the other hand, it is possible that $\cN$ be live, but not
$\transf(\cN)$. This is the case  for the net 
on the left of Figure \ref{fi-ld} (the net on the right of the same
figure is `almost' 
its Free Choice expansion).
For a detailed comparison of the  behaviors of $\cN$
and $\transf(\cN)$, 
see \cite{haar}. 

\medskip

An example of a non-Free Choice Petri net such that $\transf(\cN)$ is
live and bounded is proposed 
in Figure \ref{fi-Jean}. 

\medskip

Lemma \ref{le-bound} and \ref{le-viv} undergo the  following modifications. 

\begin{lemma}\label{le-boundviv}
Let $\pn$ be a Petri net with Free Choice expansion
$\transf(\cN)$. We have the following implications: 
\begin{center}
1. $\cN$ is bounded $\iff \ $ 2. $\transf(\cN)$ is bounded
  $\iff \ $ 3. $(\pn,u)$ is bounded for any $u$;\\ 
$a$. $\cN$ is live $\Longleftarrow \ $  $b$. $\transf(\cN)$ is live $\iff \ $
$c$. $(\pn,u)$ is live for any equitable $u$.
\end{center} 
\end{lemma}

The equivalence between $a.$ and $c.$ which was proved in Lemma \ref{le-viv}
for Free Choice nets is not true in general.
 
\begin{proof}
We have just seen that 3. implies 1.
and that 2. and 3. are equivalent. The proof of the equivalence
between 1. and 3. was done 
in Lemma \ref{le-bound}. 

Now let us prove the equivalence between $b.$ and $c$. Assume there
exists an equitable 
routing $u$ such that $(\cN,u)$ is not live. 
Construct the set $X$ of nodes of $\cN$ as in the proof of Lemma \ref{le-viv} 
(the construction there does not  require the Free Choice assumption). 
In $\transf(\cN)$,  the set $\transf(X)\cap \transf(\cP)$ is a siphon
which can be emptied 
using the same firing sequence as for $X$. 
We deduce that $\transf(X)\cap \transf(\cP)$ cannot contain an initially
marked trap, hence $\transf(\cN)$ 
cannot be live by Commoner's Theorem \ref{th:commoner}.
\end{proof}

Lemma \ref{le-boundviv} shows that the liveness and boundedness of a
routed Petri net is directly 
linked to the one of its unrouted Free Choice expansion. 

\medskip

Theorem \ref{th:blockrouted}, 
Lemma \ref{le-oneway}, Theorem \ref{th-order1}, Lemma \ref{positive}
and Proposition \ref{pr-comput} still hold when replacing the assumption 
{\em live and bounded Free Choice net} by the assumption {\em Petri
net with a live and bounded Free Choice expansion}. 
The proof of Theorem \ref{th:blockrouted}
is actually carried out below under  
the latter assumption (Theorem \ref{th-blockrouted2}). As for the
other results, it is not difficult 
to extend them by first  
considering the Free Choice expansion and then showing that the
results still hold for the original 
Petri net.

\begin{example}\label{ex-comput}\rm
Consider the live and bounded Petri net of Figure \ref{fi-Jean}. 
Clearly, it is not a Free Choice net, but its Free Choice expansion is
live and bounded.  
\begin{figure}[htb]
\begin{center}
\input{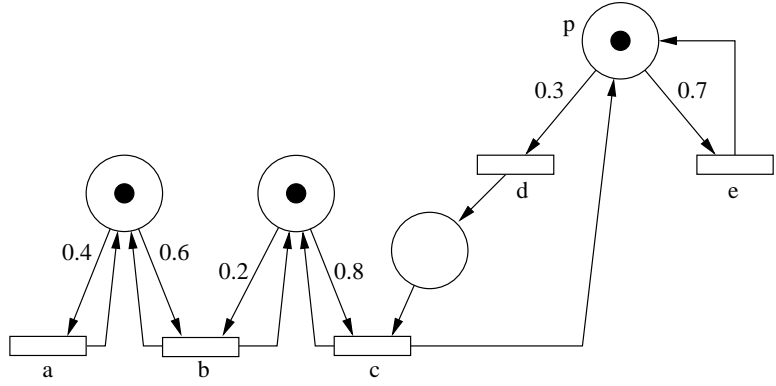}
\caption{The values on the arcs are the routing probabilities.\label{fi-Jean}}
\end{center}
\end{figure}
Consider a stochastic routed version of the Petri net. As detailed above,
the results of Theorem \ref{th-order1} and Proposition \ref{pr-comput} apply. 
In particular, let $R$ be defined as in Proposition \ref{pr-comput} and let
$\lambda=(\lambda_t, t\in \cT)$ be the vector
of throughputs (the transitions being listed in alphabetical order). 
We have

\begin{equation}\nonumber
R=\left( \begin{array}{ccccc} 0.4 & 0.3 & 0 & 0 &0 \\
                              0.4 & 0.4 & 0.4 & 0 &0 \\ 
                              0 & 0.1 & 0.4 & 0.3 &0.7 \\ 
                              0 & 0 & 0.5 & 0 &0 \\ 
                              0 & 0 & 0 & 0.3 &0.7
\end{array}\right), \ \ 
\lambda =c \left(\begin{array}{ccccc} 0.04 & 0.05 & 0.21 & 0.21 & 0.49 \end{array}\right)\:.
\end{equation}

If we assume for instance that the routing probabilities of place $p$
are  $P\{u_p(1)=d\}=x, \ P\{u_p(1)=e\}=1-x$, 
then we obtain $\lambda=c\left(2x, 3x, 12x, 12x , 12-12x\right)/(12+17x)$. 
\end{example}

To end up the section, we prove as announced a version of Theorem
\ref{th:blockrouted} for Petri nets
whose Free Choice expansion is live and bounded.

\begin{theorem}\label{th-blockrouted2}
Let $(\pn,M_0)$ be a Petri net whose Free Choice expansion
$\transf(\cN)$ is live and bounded.  
For any transition $b$, there
exists a blocking marking $M_b$  
such that for every equitable routing $u$ and all $M\in R(M_0,u)$, we have 
$R_b(M,u)=R_b'(M,u)=\{M_b\}$. 
\end{theorem}

\begin{proof}
Consider $\transf(\cN)$ and set $\cP'=\transf(\cP)-\cP$ 
and $\cT'=\transf(\cT)-\cT$,
The function $\transf$ maps a marking $M$ of $\pn$ into a marking $\transf(M)$
of $\transf(\cN)$ as defined above.
Now, we define a reverse transformation
$\psi: \N^{\transf(\cP)} \to \N^{\cP}$ which transforms a marking 
 $\tilde{M}$ of
$\transf(\cN)$
into a  marking $\psi (\tilde{M})$ of $\pn$:
\[
\psi(\tilde{M})= (\psi(\tilde{M})_p)_{p\in \cP} \ \mbox{ and }\  
\psi(\tilde{M})_p=\tilde{M}_p+\sum_{(p,q)\in \cF} \tilde{M}_{s_{pq}}\:.
\]
Note that for any marking $M$ in $\cN$, we have $\psi\circ
\transf(M)=M$.

A {\em pointed marking} $(M,f)$ of $\cN$ is a pair
formed by a marking $M$ and an assignment $f$ of each token of the
marking to an output transition.  
Formally, $f$ is an application from $\{(p,t),p\in \cP, t \in p^\bullet\}$ to $\N$,
satisfying $\sum_{t\in p^\bullet} f(p,t)=M_p$ for all place $p$. 
In $(\pn,M_0,u)$, given $M_0\move{\sigma}M'$, we denote by $(M',u,\sigma)$ the
pointed marking formed  
by $M'$ and the assignment induced by $u$ and $\sigma$: the tokens in
place $p$ are assigned  
as in \eref{eq-assi}. 
To a pointed marking $(M,f)$ of $\cN$, 
we associate the marking $\transf(M,f)$ in $\transf(\cN)$ obtained from
$\transf(M)$ by firing all 
the transitions in $\cT'$ which are compatible with the assignment. 
Note that we have $\psi\circ \transf(M,f)=M$. 
We have illustrated this in Figure \ref{fig:N_illustration}; small letters
next to a token indicate the transition to which the token is
routed.
\begin{figure}[htbp]
  \begin{center}
\input{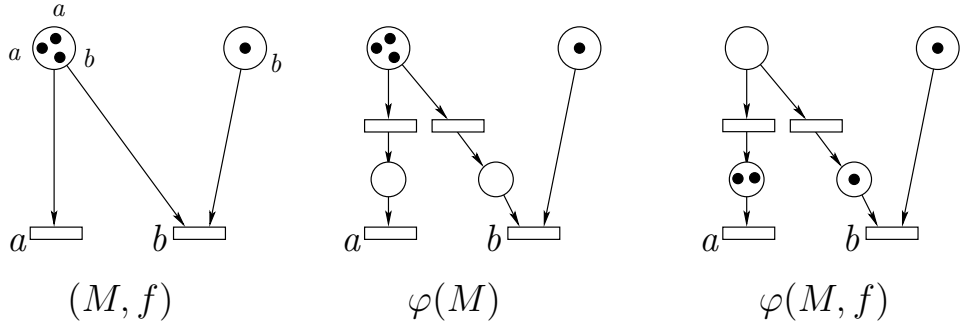}
\caption{The original net with pointed marking $(M,f)$ (left) and the effect of $\transf$.\label{fig:N_illustration}}
  \end{center}
\end{figure}

Consider the Free Choice net $\transf(\cN)$. 
By construction, any transition $b$ of $\cT$ is a non-conflicting transition 
for $\transf(\cN)$. 
Using Theorem \ref{th:want}, there exists 
a marking $M'_b$ in $\transf(\pn)$ such that for all $M\in
R(\transf(\cN),\transf(M_0))$, we have 
$R_b(\transf(\cN),M)=R_b'(\transf(\cN),M)=\{M'_b\}$. Let us
set $M_b=\psi(M_b')$.  

Consider now the routed Petri net $(\cN,M_0,u)$.
We want to prove first that $M_b$ is such that $R_b(\cN,M,u)= \{M_b\}$ 
for all $M\in R(\cN,M_0,u)$. 
Assume that there exists $M'\in R_b(\cN,M,u)$ and let $\sigma,\tau$ be such that 
$M_0\move{\sigma} M\move{\tau} M'$.
Let us consider the pointed marking $x=(M',u,\sigma\tau)$ and the marking
$\transf(x)$ of $\transf(\cN)$.  
Assume that there is a transition $t\neq b$ of $\transf(\cN)$ which is
enabled in $\transf(x)$.  
By construction, we have $t\in \cT$, and
$t$ is also enabled in
$\psi\circ \transf(x)=M'$, which is a contradiction. We conclude that $b$ is the
only transition  
enabled in $\transf(x)$, that is $\transf(x)=M_b'$, which 
implies that $M'=M_b$. 

Now we prove that $R_b'(\cN,M,u)$ is non-empty for any reachable marking $M$. 
Starting from $M$,
we build a firing sequence of the routed net by always firing an
enabled transition  
different from $b$. By Lemma \ref{le-le}, it is impossible to build an
infinite such sequence.  
Hence, we end up in a marking such  that no transition is enabled
except $b$, this marking 
belongs to 
$R_b'(\cN,M,u)$.
Since $R_b'(\cN,M,u)\subset R_b(\cN,M,u)$, this finishes the proof. 
\end{proof}

\subsection*{Acknowledgments} 
We would like to thank Javier Esparza whose suggestions greatly helped
us when we were  
blocked in our attempts to block transitions.


\end{article}

\end{document}